\documentclass[aps, pre, twocolumn, superscriptaddress, nobibnotes, nodoi, nourl, noeprint, reprint, longbibliography]{revtex4-2}

\usepackage{siunitx,amsmath,amsfonts,physics,bm}
\usepackage{xcolor,graphicx,float}
\usepackage[colorlinks=true,citecolor=blue,urlcolor=blue,linkcolor=black]{hyperref}

\newcommand{\kb}{k_{\scriptscriptstyle \rm B}}

\newcommand{\rms}{\rm \scriptscriptstyle}
\newcommand{\ri}{\vb*{r}_i}
\newcommand{\rj}{\vb*{r}_j}
\newcommand{\rjb}{\bar{\vb*{r}}_j}
\newcommand{\mub}{\vb*{\mu}^{\rms B}}
\newcommand{\murp}{\vb*{\mu}^{\rms RP}}
\newcommand{\murpb}{\vb*{\mu}^{\rms RPB}}
\newcommand{\muself}{\vb*{\mu}_{\rms self}^{\rms RPB}}
\newcommand{\murpbperp}{\mu_{\perp}^{\rms RPB}}

\begin{document}
\title{Hydrodynamic interactions hinder transport of flow-driven colloidal particles}

\author{Dominik Lips}
\affiliation{Universit\"{a}t Osnabr\"{u}ck, Fachbereich Physik, Barbarastra{\ss}e 7, D-49076 Osnabr\"uck, Germany}

\author{Eric Cereceda-L\'{o}pez}
\affiliation{Departament de F\'{i}sica de la Mat\`{e}ria Condensada, Universitat de Barcelona, 08028, Barcelona, Spain}
\affiliation{Institut de Nanoci\`{e}ncia i Nanotecnologia, Universitat de Barcelona (IN2UB), 08028, Barcelona, Spain}

\author{Antonio Ortiz-Ambriz} 
\affiliation{Departament de F\'{i}sica de la Mat\`{e}ria Condensada, Universitat de Barcelona, 08028, Barcelona, Spain}
\affiliation{Institut de Nanoci\`{e}ncia i Nanotecnologia, Universitat de Barcelona (IN2UB), 08028, Barcelona, Spain}
\affiliation{University of Barcelona Institute of Complex Systems (UBICS), 08028, Barcelona, Spain}
\affiliation{Tecnol\'ogico de Monterrey, Escuela de Ingenier\'ia y Ciencias, Campus Monterrey, 64849, Monterrey, Mexico}

\author{Pietro Tierno} 
\affiliation{Departament de F\'{i}sica de la Mat\`{e}ria Condensada, Universitat de Barcelona, 08028, Barcelona, Spain}
\affiliation{Institut de Nanoci\`{e}ncia i Nanotecnologia, Universitat de Barcelona (IN2UB), 08028, Barcelona, Spain}
\affiliation{University of Barcelona Institute of Complex Systems (UBICS), 08028, Barcelona, Spain}

\author{Artem Ryabov}
\affiliation{Charles University, Faculty of Mathematics and Physics, Department of Macromolecular Physics, V Hole\v{s}ovi\v{c}k\'{a}ch 2, 
CZ-18000 Praha 8, Czech Republic}

\author{Philipp Maass} 
\email{maass@uos.de}
\affiliation{Universit\"{a}t Osnabr\"{u}ck, Fachbereich Physik, Barbarastra{\ss}e 7, D-49076 Osnabr\"uck, Germany}

\date{November 15, 2022} 

\begin{abstract}
The flow-driven transport of interacting micron-sized particles occurs in many soft matter systems spanning from the translocation of proteins to moving emulsions in microfluidic devices. Here we combine experiments and theory to investigate the collective transport properties of colloidal particles along a rotating ring of optical traps. In the corotating reference frame, the particles are driven by a vortex flow of the surrounding fluid. When increasing the depth of the optical potential, we observe a jamming behavior that manifests itself in a strong reduction of the current with increasing particle density. We show that this jamming is caused by hydrodynamic interactions that enhance the energetic barriers between the optical traps. This leads to a transition from an over- to an under-critical tilting of the potential in the  corotating frame. Based on analytical considerations, the enhancement effect is estimated to increase with increasing particle size or decreasing radius of the ring of traps. Measurements for different ring radii and Stokesian dynamics simulations for corresponding particle sizes confirm this. The enhancement of potential barriers in the flow-driven system is contrasted to the reduction of barriers in a force-driven one. This diverse behavior demonstrates that hydrodynamic interactions can have a very different impact on the collective dynamics of many-body systems. Applications to soft matter and biological systems require careful consideration of the driving mechanism and of the role of hydrodynamic interactions.
\end{abstract}
\maketitle

\section{Introduction}
\label{sec:introduction}
At the nano and micrometer scale, the dynamics of colloidal particles dispersed in a fluid occur at low Reynolds numbers,
where viscous friction dominates over inertia. At high densities, 
the particles interact via long-range hydrodynamic interactions (HIs), which are mediated by the flow of the dispersing medium \cite{Happel/Brenner:1973, Dhont:1996, Hess/Klein:2006}.
These interactions play an important role
in many biological and soft matter systems \cite{Lauga/Powers:2009}. For example, 
they influence the organization of sperm cells \cite{Riedel/etal:2005} and of magnetotactic bacteria \cite{Pierce/etal:2018},
modify the dynamics of bacteria propelling close to a surface \cite{Lauga/etal:2006, Berke/etal:2008, Leonardo/etal:2011}
or they induce synchronization phenomena \cite{Reichert/Stark:2005, Vilfan/Juelicher:2006, Drescher/etal:2009, Uchida/Golestanian:2011, Brumley/etal:2012, Chakrabarti/Saintillan:2019}.

Contrary to biological organisms such as bacteria exhibiting non-trivial shapes and interactions,
spherical colloidal particles are a relatively simple model system allowing
to investigate the impact of HIs on transport properties.
Since the particle size is within the visible wavelength, digital video microscopy \cite{Crocker1996,Baumgartl2005} can be used to extract the particle trajectories, and 
measure how the dispersing medium influences the collective dynamics. 
In particular, the impact of HIs in 
passive, i.e. diffusing, colloidal particles has been the subject of long research to date \cite{Qiu/etal:1990, Zahn/etal:1997, Naegele/Baur:1997, Rinn/etal:1999, Haertl/etal:2000, Riese/etal:2000, Santana/etal:2001, Banchio/etal:2006}.
The effect of HIs has also been investigated in the context of 
particle sedimentation \cite{Segre/etal:1997, Segre/etal:2001, Padding/Louis:2004}, confinement \cite{Cui/etal:2002, Cui/etal:2004, Xu/etal:2005}, pattern formation \cite{Grzybowski/etal:2000, Lenz/etal:2003} or 
crystallization kinetics \cite{Radu/Schilling:2014, Tateno/etal:2019}.

\begin{figure*}[t!]
\centering
\includegraphics[width=\textwidth]{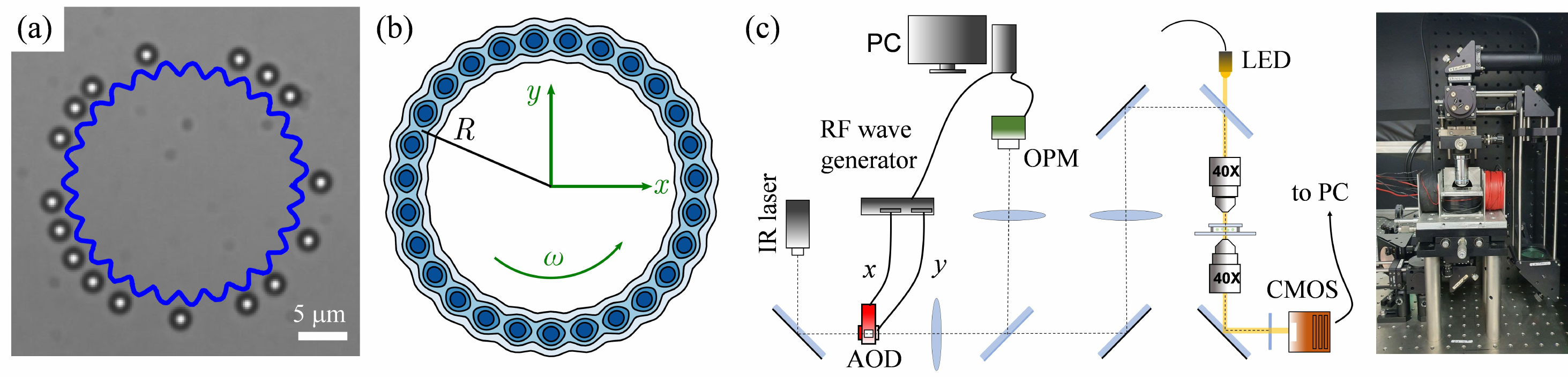}
\caption{Illustration of the experimental setup. (a) Optical microscope image showing $N=17$ colloidal particles 
of radius $a=2\,\si{\mu m}$ trapped along a ring with 
$N_{\rm tr}=27$ equidistant optical traps (blue line). 
(b) Isolines of the periodic potential generated by the scanning laser 
tweezers with angular velocity $\omega$. The experiments are performed for
different ring radii  $R=17.8\,\si{\mu m}$, $20.2\,\si{\mu m}$, and $21.8\,\si{\mu m}$.
(c) Schematic of the experimental setup to generate 
the ring of optical traps. A continuous-wave infrared laser is passed through 
a two-channel ($x,y$) acousto-optic deflector (AOD).
The beam is divided such that a fraction goes  
through an optical power meter (OPM). This is used to keep the 
beam intensity constant via a feedback loop.
The beam is focused by a microscope objective ($40\times$ magnification) to the sample cell. 
Visualization of the particles is obtained with a light-emitting diode (LED) 
coupled to a Complementary Metal-Oxide-Semiconductor (CMOS) camera.
In a reference frame corotating with the traps, an ideal periodic optical potential would be
time-independent and the particles are driven by a vortex flow, see Eq.~\eqref{eq:langevin-corotating}.}
\label{fig:experimental_setup}
\end{figure*}

In driven systems, when external forces
are used to induce particle motion, 
HIs may profoundly affect the collective dynamics 
giving rise to unexpected effects \cite{Zahn/etal:1997, Rinn/etal:1999, Reichert/Stark:2004, Lutz/etal:2006, Beatus/etal:2007, Beatus/etal:2009, Grimm/Stark:2011, Malgaretti/etal:2012, Dobnikar/etal:2013, Nagar/Roichman:2014, Martinez/Tierno:2018, Misiunas/Keyser:2019}.
For example, electrophoretically driven particles confined in a narrow, microfluidic channel display a 
speed-up effect at high densities \cite{Misiunas/Keyser:2019}.
In this context, optical tweezers represent powerful tools 
to trap particles and measure 
the effect of HIs between pairs \cite{Meiners/Quake:1999, Dufresne/etal:2000, Martin/etal:2006, Kotar/etal:2013, Kreiserman/etal:2019} or denser assemblies \cite{Ladavac/Grier:2005, Polin/etal:2006}. 

While the impact of HIs on particle transport has been studied in detail for force-driven systems \cite{Reichert/Stark:2004, Lutz/etal:2006},
the impact on flow-driven systems has not been much investigated yet. These systems are important also, 
for example, when considering the blood flow in organisms or the flow in microfluidic devices.

In this article we combine experiments and numerical simulations to investigate a flow-driven system. 
In the experiments, micron-sized polystyrene particles are confined along a ring 
via time-shared optical tweezers, which create a controlled number of optical potential wells. These traps are rotated
at a fixed angular frequency and drag the particles along with them. In the co-rotating reference frame, each particle is 
driven by a vortex flow field. This can be understood by considering an observer at the ring center rotating clockwise with the optical traps. For this observer, the traps would stay at fixed positions and the fluid would flow counterclockwise, corresponding to a vortex flow.
In addition, the motion of one particle is influenced by the flow fields generated by 
the motion of the other particles, i.e.\ by the HIs. 

To explore the impact of the HIs in this system,
we perform Stokesian dynamics simulations, where the HIs are taken into account in the Langevin equations with
mobility tensors depending on the position of all particles. These simulations allow us to understand
the experimentally observed slowdown of the particle transport with increasing density for deep optical traps, 
which reflects a jamming behavior. This work aims at extending previous results  \cite{Cerendeca-Lopez:2021} by providing a more detailed description of our findings and explanations. Further, we experimentally analyze different situations by varying both the radius of the optical ring and the strength of the potential wells. We 
provide also more details on the experimental setup, a description on how the time-dependent traveling wave like potential in the experiment was extracted from video-recorded single-particle trajectories.
Based on the mobility method for HIs \cite{Kim/Karrila:1991}, we also explore the
impact of the HI when instead of the flow driving, a force driving of analogous type and with equal strength is present.
This allows us to compare the impact of HIs on flow- and force-driven systems. Remarkably, we find huge differences. 
While under flow-driving, the particle transport is mitigated, it is facilitated under force driving. We discuss the origins 
of this strikingly different behavior.

The manuscript is organized as follows. 
In section~\ref{sec:experimental-setup} we give a detailed description of the experimental setup. In 
Sec.~\ref{sec:potential-determination} we show that the arrangement of optical traps along the ring generates a sinusoidal potential along the azimuthal direction. We describe a procedure for extracting the real potential shape from the experimental system. Application of this procedure yields a slight modulation of the sinusoidal potential. 
In Sec.~\ref{sec:simulations}, we explain why the dynamics in the corotating frame corresponds to that in a flow-driven system [see Eq.~\eqref{eq:langevin-corotating}]. We also describe the details of the Stokesian dynamics simulations.
The impact of HIs on the collective transport properties is discussed in Sec.~\ref{sec:current-density-relations}.
In addition we compare the particle
transport under flow-driving with that under a corresponding force driving and analyze the origin of strong differences between these two forcing mechanisms in Sec.~\ref{sec:flow-vs-force-driving}. Details of our implementation of HI are given in the Appendix. 

\section{Experimental Setup}
\label{sec:experimental-setup}
As shown in Figs.~\ref{fig:experimental_setup}(a), \ref{fig:experimental_setup}(b),
we experimentally realize a flow-driven model system
by using $N$ colloidal particles trapped in a circular array of $N_{\rm tr}=27$ optical traps.  The traps are created by a continous-wave 
infrared laser beam (manlight ML5-CW-P/TKS-OTS) with wavelength $1064\,\si{\mu m}$ operating at a power of $P=3\,\si{W}$. The 
laser light passes through two acousto-optic deflectors (AOD) with two channels (AA Optoelectronics DTSXY-400-1064), which deflect 
the beam and their plane is conjugated to the back focal plane of a trapping microscope objective 
(40$\times$ Nikon water immersion, operating dry) such that it produces a trap at a position 
roughly proportional to the input frequency.
The input frequency of the AOD ranges from $60$ to $90\,\si{MHz}$ 
and it is generated by a two-channel radio frequency wave generator (DDSPA2X-D431b-34). 
The wave generator in turn is addressed by a digital output card (National Instruments cDAQ NI-9403) 
with a refresh frequency of $150\,\si{kHz}$. This card inputs a value of frequency to the wave generator at 
$50\,\si{kHz}$ (3 steps are required: lock, set, unlock), 
and therefore, the AOD is capable of deflecting the beam to a different position every $20\,\si{\mu s}$. 
For $27$ traps, this means that each position is visited once every $\sim 0.5\,\si{ms}$, which is much faster than the self-diffusion time 
$\approx30\,\si{s}$ of the particles, effectively producing simultaneous traps. 

A detailed scheme of the experimental setup with the principal optical components, 
and a small image as an inset is shown in Fig.~\ref{fig:experimental_setup}(c). 
A conjugating pair of lenses is used to make all
beams converging to the back focal plane of the microscope objective.
A constant fraction of the light is redirected 
using a different set of conjugating
pair lenses to an optical power meter.
The latter is directly connected to the laser source via 
a feedback loop, which allows to monitor 
the output power in order to keep 
it constant.
 
To build a sample, we place polystyrene beads of radius $a=2\,\si{\mu m}$ (CML Molecular Probes) 
in a closed fluid cell, realized with two coverslips, and sealed with parafilm and a vacuum grease.
The sample is placed on a custom-built inverted optical microscope, equipped with a second observation objective 
(Nikon $40\times$ plan apo) and a CMOS camera (Ximea MQ003MG-CM) to record video at $30\,\si{fps}$. 
White light emitted from a LED, see Fig.~\ref{fig:experimental_setup}(c),
is sent to the experimental chamber 
passing through a dichroic mirror (DMSP950R Thorlabs, $950\,\si{nm}$ cutoff wavelength).
Video recording and the AOD driving are all done with free LabVIEW software available from GitHub. 

The advantage of our circular geometry is that it mimics periodic boundary conditions. However, applying a flow to drive the particles 
over the circular potential would require a vortex generated at the center of the ring. We found a much simpler solution,
which is to rotate the potential at a constant uniform angular velocity $\omega$. This approach differs from the 
previously described force-driving \cite{nagar_collective_2014, sassa_hydrodynamically_2012a}. 
Particles are not subjected to a constant optical force, but to a time-dependent torque, 
as detailed in Sec.~\ref{sec:potential-determination} [see Eq.~\eqref{eq:tauopt}].

\section{Optical potential landscape}
\label{sec:potential-determination}

\subsection{Expected shape of the potential}
\label{subsec:potential-shape}
The potential of a single optical trap positioned at $(x_{\rm tr}, y_{\rm tr})$ in Fig.~\ref{fig:experimental_setup}(b) 
can be described by the Gaussian profile
\begin{equation}
U_{\rm tr}(x,y;x_{\rm tr},y_{\rm tr}) = -u_0 \exp[-\frac{(x-x_{\rm tr})^2}{2w^2} -\frac{(y-y_{\rm tr})^2}{2w^2}] \, ,
\end{equation}
where $w=2.02\,\si{\mu m}$ is the width of a trap and $u_0$ its depth that is controlled by the laser power. 
For describing the circular arrangement of the traps in Figs.~\ref{fig:experimental_setup}(a) and (b), we use polar coordinates, where
$(x,y)=(r\cos\varphi,r\sin\varphi)$ and the trap center positions are at 
$(x_k,y_k)=(R\cos\varphi_k,R\sin\varphi_k)$, $k=1,\ldots,N_{\rm tr}$ with $\varphi_k=2\pi k/ N_{\rm tr}$.
Here $R$ is the ring radius.

The static potential of all optical traps is 
\begin{align}
U^{\rm st}(r, \varphi)&= \sum_{k=1}^{N_{\rm tr}} U_{\rm tr}(x,y;x_k,y_k)\label{eq:Ust-1}\\
& \hspace{-5ex}=-u_0\sum_{k=1}^{N_{\rm tr}} \exp \left[-\frac{r^2 + R^2 -2rR\cos (\varphi - \varphi_k)}{2 w^2}\right].
\nonumber
\end{align}
Due to the strong radial confinement of the particles, displacements of the colloidal particles from $r=R$ in the radial directions are almost negligible.\footnote{Videos demonstrating the negligible particle motion in radial direction can be found at 
\href{http://link.aps.org/supplemental/10.1103/PhysRevLett.127.214501}{doi:10.1103/PhysRevLett.127.214501}.}\\

Hence, we can set $r =R$, leading to the potential
\begin{align}
U^{\rm st}_\varphi(\varphi)=-u_0\sum_{k=1}^{N_{\rm tr}} \exp[-\frac{R^2}{w^2} \left(1-\cos (\varphi - \varphi_k) \right)]
\label{eq:Ust-2}
\end{align}
that depends on the azimuthal angle $\varphi$ only. 

This potential has the period $2\pi/N_{\rm tr}$,
$U^{\rm st}_\varphi(\varphi+2\pi/N_{\rm tr})=U^{\rm st}_\varphi(\varphi)$. Expanding it in a Fourier series, we obtain
\begin{equation}
U^{\rm st}_\varphi(\varphi) = \sum_{n=-\infty}^{\infty} c_n \exp(iN_{\rm tr}n\varphi)\,,
\end{equation}
where
\begin{align}
c_n&=\frac{N_{\rm tr}}{2\pi}\int\limits_{-\pi/N_{\rm tr}}^{\pi/N_{\rm tr}} 
\hspace{-0.4em}\dd\varphi\, U^{\rm st}_\varphi(\varphi)  \exp(-iN_{\rm tr}n\varphi)\label{eq:cn1}\\
&\hspace{-2ex} \cong -u_0\frac{N_{\rm tr}}{2\pi} 
\hspace{-0.5em}\int\limits_{-\pi/N_{\rm tr}}^{\pi/N_{\rm tr}} 
\hspace{-0.6em}\dd\varphi 
 \exp[-\frac{R^2}{w^2} \left(1-\cos\varphi \right)]\exp(-iN_{\rm tr}n\varphi).
\nonumber
\end{align}
In going from the first to the second line in Eq.~\eqref{eq:cn1}, we have kept only the summand for $k=N_{\rm tr}$
when inserting $U^{\rm st}_\varphi(\varphi)$ from Eq.~\eqref{eq:Ust-2}. This is allowed because of the large ratio 
$(R^2/w^2)\simeq 100$ in the experiments, 
which means that the summands for $k\ne N_{\rm tr}$ are negligible in the integration interval.
Likewise, we can set $(1-\cos\varphi)\cong \varphi^2/2$ in Eq.~\eqref{eq:cn1} and extend the limits of the 
integration to $\pm\infty$. The resulting Gaussian integral gives
\begin{align}
c_n\cong -u_0\frac{N_{\rm tr} w}{\sqrt{2\pi}R} \exp[-\left(\frac{N_{\rm tr} w}{R}\right)^2 \frac{n^2}{2}] \, .
\end{align}
Since $(N_{\rm tr} w/R)^2/2\cong3.6$, the Fourier coefficients decay quickly. The ratio $c_2/c_1$ is already of the order $10^{-5}$, which allows a truncation of the Fourier series after $n=1$, leading to a sinusoidal form of 
\begin{equation}
U^{\rm st}_\varphi(\varphi)\cong c_0+\frac{U_0}{2}\cos( N_{\rm tr} \varphi)
\label{eq:potential-static}
\end{equation}
with potential barrier
\begin{equation}
U_0=4c_1=-2u_0\sqrt{\frac{2}{\pi}}\frac{N_{\rm tr}  w}{R} \exp[-\frac{1}{2}\left(\frac{N_{\rm tr} w}{R}\right)^2]\,.
\label{eq:A0}
\end{equation}
A similar derivation of this result has been given in Ref.~\cite{Juniper/etal:2016}, 
and a sinusoidal form of the potential resulting from a periodic arrangement of Gaussian optical traps has been seen before also in
the analysis of experiments \cite{Juniper/etal:2012}.
For the radii $R$ studied in our experiments, the relative deviation between the $U^{\rm st}_\varphi(\varphi)$ from 
Eq.~\eqref{eq:Ust-2} and the approximate sinusoidal
potential \eqref{eq:potential-static} is less than 0.15\% for all $\varphi$. 

The rotation of the traps along the ring with angular frequency $\omega$ leads to the time-dependent potential
\begin{equation}
U_\varphi(\varphi,t)=U^{\rm st}_\varphi(\varphi-\omega t)=c_0+\frac{U_0}{2}\cos(N_{\rm tr} (\varphi-\omega t))
\label{eq:Uexpected-laboratory}
\end{equation}
in the laboratory frame. In a corotating reference frame, this traveling-wave type potential reduces to 
$U^{\rm st}_\varphi(\varphi)$.

\subsection{Extracting potential shape from experimental data}
\label{subsec:potential-shape-procedure}
In experiments, the potential can be evaluated by analyzing the single-particle dynamics. 
For the kinematic viscosity $\nu\simeq 10^{-6}\,\si{m^2/s}$ of water and azimuthal velocity $\omega R$, the Reynolds
number is ${\rm Re}=\omega R\sigma/\nu\simeq 3\times10^{-5}$, i.e.\ much smaller than one.
The velocity correlation time $(4\pi a^3 \rho_{\rms PS}/3)/(6\pi \rho_{{\rms H}_2{\rms O}}\nu a)\simeq 1\si{\mu s}$ 
for the polystyrene beads is much smaller than
the characteristic diffusion time $a^2/D_0\simeq30\,\si{s}$ ($\rho_{\rms PS}$, 
$\rho_{{\rms H}_2{\rms O}}$: densities of polystyrene and water). Under these conditions, inertia effects can be neglected
and a single-particle performs an overdamped Brownian motion described by the Langevin equation
\begin{equation}
\frac{\dd{\bm r}}{\dd t}=-\mu_0\bm\nabla U_{\rm opt}(\bm r,t)+\bm\zeta(t)\,,
\label{eq:langevin-1}
\end{equation}
where $U_{\rm opt}(\bm r,t)$ is the potential of the optical forces,
$\mu_0$ is the mobility, and $\bm\zeta(t)$ is a Gaussian white noise with $\langle\bm\zeta(t)\rangle=0$ and 
$\langle\zeta_\alpha(t)\zeta_\beta(t')\rangle=2D_0\delta_{\alpha\beta}\delta(t-t')$; $D_0=k_{\rm B}T\mu_0$ is the diffusion coefficient.

As we explained after Eq.~\eqref{eq:Ust-1},  we need to consider
particle displacements in the azimuthal direction only. The Langevin equation~\eqref{eq:langevin-1} 
then simplifies to
\begin{equation}
\frac{\dd\varphi}{\dd t}
=\frac{\mu_0}{R^2}\,\tau_{\rm opt}(\varphi,t)+\zeta_\varphi(t)\,,
\label{eq:dvarphi-lab}
\end{equation}
where $\langle \zeta_\varphi(t)\rangle=0$,
$\langle \zeta_\varphi(t)\zeta_\varphi(t')\rangle=2D_0R^{-2}\delta(t-t')$, and
\begin{equation}
\tau_{\rm opt}(\varphi,t)=-\frac{\partial U_\varphi(\varphi,t)}{\partial\varphi}
\label{eq:tauopt}
\end{equation}
is the torque excerted on the particle. According to Eq.~\eqref{eq:Uexpected-laboratory}, this is expected to
be a travelling wave $(N_{\rm tr}U_0/2)\sin(N_{\rm tr}(\varphi-\omega t))$ in the laboratory frame.

In the corotating frame with the angle variable $\varphi'=\varphi-\omega t$, Eq.~\eqref{eq:dvarphi-lab} becomes
\begin{equation}
\frac{\dd\varphi'}{\dd t}
=\frac{\mu_0}{R^2}\,\left[-\frac{\omega R^2}{\mu_0}+\tau'_{\rm opt}(\varphi')\right]+\zeta_\varphi(t)\,,
\label{eq:dvarphi-corot}
\end{equation}
where 
$\tau'_{\rm opt}(\varphi')=(N_{\rm tr}U_0/2)\sin(N_{\rm tr}\varphi')$
is a time-independent torque. This would be the ideal mathematical form, but one cannot
expect the experimental setup to generate this in a perfect manner. 

The real torque in the corotating frame can have a time dependence,
$\tau'_{\rm opt}=\tau'_{\rm opt}(\varphi',t)$, which for our experimental setup must be periodic,
\begin{equation}
\tau'_{\rm opt}\left(\varphi',t+\frac{2\pi}{\omega}\right)=\tau'_{\rm opt}(\varphi',t)\,.
\end{equation}

To determine $\tau'_{\rm opt}(\varphi',t)$, we divide the periodicity intervals $[0,2\pi[$ and $[0,2\pi/\omega[$ of $\tau'_{\rm opt}(\varphi',t)$ into
$N_\varphi=270$ (ten per wavelength of the optical potential)
and $N_\omega=15$ bins of identical widths $\Delta\varphi'=2\pi/N_\varphi$ and $\Delta t=2\pi/N_\omega\omega$, respectively. 
These numbers ensure that we obtain a good resolution along the azimuthal direction and in time, while
having sufficient statistical accuracy.
The bin intervals are $\mathfrak{J}_j=[(j-1)\Delta\varphi',j\Delta\varphi'[$, $j=1,\ldots,N_\varphi$, and $\mathcal{K}_k=[(k-1)\Delta t,k\Delta t[$, $k=1,\ldots,N_\omega$ and 
have the midpoints $\varphi'_j=(2j-1)\Delta\varphi'/2$ and $t_k=(2k-1)\Delta t$. The torque
$\tau'_{\rm opt}(\varphi',t)$ is obtained by the
first Kramers-Moyal coefficient \cite{Risken:1985},
\begin{align}
\tau'_{\rm opt}(\varphi'_j,t_k)&=\label{eq:kramers-moyal}\\
&\hspace{-3.5em}\frac{R^2}{\mu_0}\left[
\frac{\Bigl\langle [\varphi'(t+\delta t)-\varphi'(t)]\, \bigl |\, \varphi'(t)\in \frak{J}_j,\, t\in\mathcal{K}_k\Bigr\rangle}{\delta t}-\omega\right],
\nonumber
\end{align}
where $\varphi'(t)$ is a single-particle trajectory
in the corotating frame, measured with a time resolution $\delta t$. The $\langle\ldots\rangle$ means an average over many times in the measurement under the condition that $\varphi'(t)$ is in the bin interval $\frak{J}_j$ and the time $t$ in the bin interval $\mathcal{K}_k$.
When applying Eq.~\eqref{eq:kramers-moyal}, the angle $\varphi=\varphi(t)$ and the time $t$ are first mapped to the intervals
$[0,2\pi[$ and $[0,2\pi/\omega[$.

\begin{figure*}[t!]
\centering
\includegraphics[width=\textwidth]{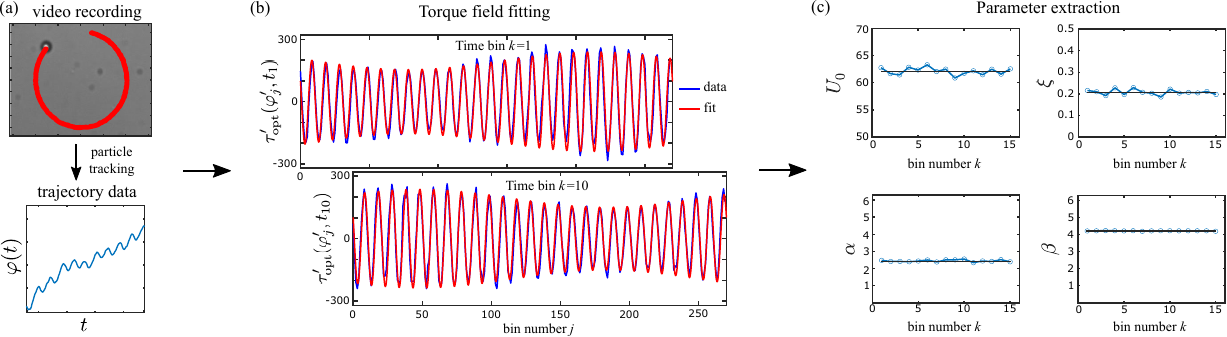}
\caption{Method for determining potential parameters from single-particle measurements.
(a) Illustration of the particle tracking. (b) Fitting of the torque. The blue lines represent
representative results for the torque $\tau'_{\rm opt}(\varphi',t)$ at two different times 
$t_1$ and $t_{10}$ in the corotating frame obtained from measured trajectories using Eq.~\eqref{eq:kramers-moyal}. The red lines are fits of Eq.~\eqref{eq:Ureal_corotating-2} to the experimental data. 
(c) Parameters $U_0$, $\xi$, $\alpha$ and $\beta$ extracted from the fitting.}
\label{fig:potential_parameters}
\end{figure*}

\subsection{Real potential}
\label{subsec:real-potential}
In the experiments, we placed a single particle on the ring and recorded
a video for 30~min at a time resolution of 30~frames per second ($\delta t=1/30\,\si{s}$). The video is analyzed by a particle tracking algorithm that outputs a two-dimensional trajectory. After conversion to polar coordinates, we obtain two time series $r(t)$ and $\varphi(t)$. Because of the strong radial confinement, we proceed by only analyzing $\varphi(t)$. 
After transforming $\varphi(t)$ to the corotating frame,
\begin{equation}
\varphi'(t)=\varphi(t)-\omega t\,,
\end{equation}
we determined $\tau'_{\rm opt}(\varphi_j,t_k)$ from Eq.~\eqref{eq:kramers-moyal}.

The mobility $\mu_0=D_0/k_{\rm B}T$ in Eq.~\eqref{eq:kramers-moyal} was obtained independently
from measurements of the time-dependent mean-square displacements in the absence of the optical force fields, which
yielded the diffusion coefficient $D_0\cong 0.1295\,\si{\mu m^2s^{-1}}$.

Figure~\ref{fig:potential_parameters}(a) illustrates the particle tracking and Figure~\ref{fig:potential_parameters}(b) 
shows representative results 
for the torque at two different times in the corotating frame, see the 
blue lines in the graph for $\tau'_{\rm opt}(\varphi_j,t_1)$ and $\tau'_{\rm opt}(\varphi_j,t_{10})$. 
The data show a deviation from the ideal travelling  wave form, i.e.\ the sinus function for the torque in the corotating frame. 

This deviation is likely caused by a non-flat amplitude response of the AOD in the deflection space. 
Such an imperfection can destroy the $(2\pi/N_{\rm tr})$-periodicity of the optical potential but the potential
must still be $(2\pi)$-periodic with respect to $\varphi$. 
Considering the potential barrier to be a general $(2\pi)$-periodic function of $\varphi$, we may take the 
first term of a Fourier expansion to describe the modified potential shape. This gives
\begin{equation}
U_\varphi(\varphi, t) = \frac{U_0}{2} \left[1 + \xi \sin(\varphi)\right] \cos \left( N_{\rm tr} (\varphi - \omega t) \right)
\label{eq:Ureal_laboratory}
\end{equation}
in the laboratory frame, and 
\begin{equation}
U'_\varphi(\varphi', t) = \frac{U_0}{2} \left[1 + \xi \sin(\varphi' + \omega t)\right] \cos \left( N_{\rm tr} \varphi' \right)
\label{eq:Ureal_corotating}
\end{equation}
in the corotating frame. According to Eq.~\eqref{eq:Ureal_laboratory}, the barrier height is modulated with a strength $\xi$, and $U_0$ thus represents a mean barrier height.

For comparison with the torque data in Fig.~\ref{fig:potential_parameters}, we must consider phase shifts 
$\alpha$ and $\beta$ for both the 
periodic trap arrangement and the amplitude modulation. Taking the time instant $t=0$, the shift 
$\alpha$ accounts for the phase difference between the
maxima of the $\cos(N_{\rm tr}\varphi)$ function and an arbitrary but fixed zero point of $\varphi$ along the ring in 
Fig.~\ref{fig:experimental_setup}(b). Likewise, the shift $\beta$ accounts for the phase difference between the
maximum of the barrier height modulation and the zero point. In the corotating frame we then have
\begin{equation}
U'_\varphi(\varphi', t)
=\frac{U_0}{2} \left[1 + \xi \sin(\varphi' + \beta+\omega t)\right] \cos\left( N_{\rm tr} (\varphi'+\alpha) \right)
\label{eq:Ureal_corotating-1}
\end{equation}
and
\begin{align}
\tau'_{\rm opt}(\varphi',t)=&\frac{U_0}{2} \Bigl\{\xi\cos(\varphi' + \beta+\omega t)\cos\left( N_{\rm tr} (\varphi'+\alpha) \right)
\label{eq:Ureal_corotating-2}\\
&\hspace{-2em}{}+N_{\rm tr}\left[1 + \xi \sin(\varphi' + \beta+\omega t)\right] 
\sin\left( N_{\rm tr} (\varphi'+\alpha) \right)\!\Bigr\}\,.\nonumber
\end{align}
To determine the yet unknown parameters $U_0$, $\xi$, $\alpha$ and $\beta$, we
fitted Eq.~\eqref{eq:Ureal_corotating-2} to the experimental data for each time bin. The results are shown in 
Fig.~\ref{fig:potential_parameters}(c). For all parameters we obtain constant values up to very small numerical 
uncertainties, which demonstrates that the functional form \eqref{eq:Ureal_corotating-1} is accurately representing the 
optical potential.

\begin{figure}[b!]
\centering
\includegraphics[scale=1]{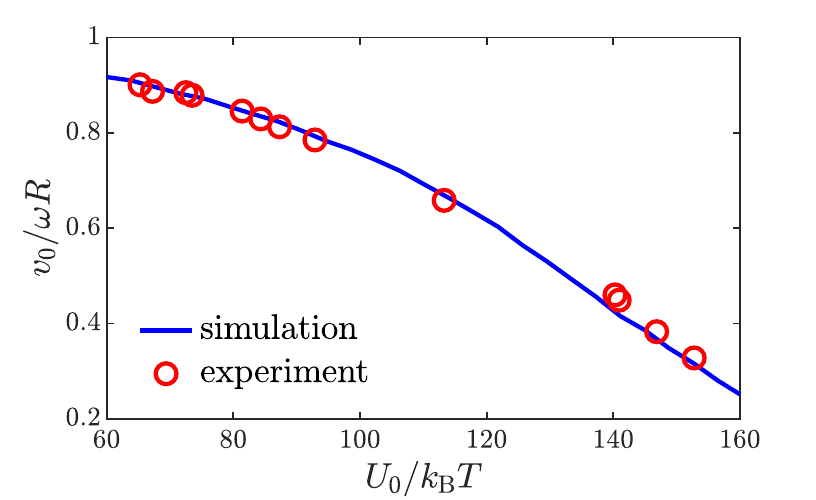}
\caption{Comparison of measured (symbols) with simulated (line) mean velocity of a single particle as a function of the mean barrier height $U_0$
for ring radius $R=20.2\,\si{\mu m}$, angular frequency $\omega=0.63\,\si{rad/s}$
of optical trap rotation, and potential amplitude modulation $\xi=0.22$ 
[see Eqs.~\eqref{eq:Ureal_laboratory}, \eqref{eq:Ureal_corotating}].}
\label{fig:v0}
\end{figure}

We critically checked the procedure of determining the optical potential by generating 
simulated data replacing the experimental ones. The corresponding simulations were performed for the Brownian dynamics of a single particle in the potential \eqref{eq:Ureal_corotating-1} with the parameters  estimated from the measured trajectories in Fig.~\ref{fig:potential_parameters}(c).
Particle trajectories in these simulations were recorded with the same time resolution $\delta t$ as in the experiment, and the procedure described above was applied to evaluate the potential parameters. 

We found that all parameters agreed with the ones estimated from the experimental trajectories except of $U_0$, which was underestimated by a few precent. The reason for this underestimation is that the time resolution in the experiment was not fully sufficient ($\delta t$ a bit too large). When using a higher time resolution in the generation of the data, the procedure for determining the potential parameters was giving the correct value for $U_0$ also. Figure~\ref{fig:v0} shows that the simulated $v_0$ (solid line) agree well with the measured ones (symbols). 

\section{Stokesian dynamics simulations}
\label{sec:simulations}
The driven Brownian motion of the $N$ colloidal particles can be described by the Langevin equations
\begin{equation}
\frac{\dd\bm{r}_i}{\dd t}=\sum_{j=1}^{N}
\bigl[k_{\rm B}T\bm\nabla_j \bm\mu_{ij}-\bm\mu_{ij}\bm\nabla_j U_{\rm opt}(\bm{r}_j,t)\bigr] + \bm\zeta_i\,,
\label{eq:langevin-laboratory}
\end{equation}
where $\bm\zeta_i$ is a vector describing Gaussian white noise processes with
zero mean and covariance matrix in accordance with the fluctuation-dissipation theorem, i.e.\
$\langle\bm\zeta_i(t)\otimes\bm\zeta_j(t')\rangle=2k_{\rm B} T\bm\mu_{ij}\delta(t-t')$. 
The mobility tensor $\bm\mu_{ij}$ describes the force on particle $i$ mediated by the flow fields generated by the 
other particles $j$ as well as flow-effects mediated by the coverslip surface. It depends on the positions of the 
other particles and the distance from the surface. 

For point particles and a flat surface with no-slip boundary conditions, 
$\bm\mu_{ij}$ is given by the Blake tensor \cite{Blake:1971}. The entrainment of the fluid due to the
finite size of the colloidal particles can be accounted for by a multipole expansion. Its truncation at the second
order yields the Rotne-Prager level of the Blake tensor \cite{Hansen/etal:2011}. 
The respective formulas for $\bm\mu_{ij}$ are given in the Appendix.

In a reference frame corotating with the optical traps, the particles are dragged by the surrounding fluid and their Brownian dynamics become flow-driven. 
To see this, let us transform Eq.~\eqref{eq:langevin-laboratory} into the corotating frame, where we denote the particle positions by $\bm r^{\,\prime}_i$. The transformation of the positions between laboratory and comoving frame is given by
$\bm r_i=\bm r^{\,\prime}_i+\bm\omega\times\bm r^{\,\prime}_i\, t$. Accordingly, we obtain
\begin{align}
\frac{\dd {\bm{r}^{\,\prime}_i}}{\dd t}=&-\bm\omega\times\bm r^{\,\prime}_i\label{eq:langevin-corotating}\\
&{}+\sum_{j=1}^{N}
\bigl[k_{\rm B}T\bm\nabla^{\,\prime}_j \bm\mu_{ij}'-\bm\mu_{ij}'\bm\nabla^{\,\prime}_j U_{\rm opt}'(\bm{r}^{\,\prime}_j,t)\bigr]
+ \bm\zeta_i(t)\,,
\nonumber
\end{align}
where the prime in $\bm\nabla^{\,\prime}_i$ means to take the derivative with respect to the primed coordinates, 
$U_{\rm opt}'(\bm{r}^{\,\prime}_i,t)=U_{\rm opt}(\bm{r}_i-\bm\omega\times\bm r_i\,t,t)$, and $\mu_{ij}'$ are the elements of the mobility tensors as functions of the primed coordinates in the corotating frame. The first term on the right hand side of the equations corresponds to a vortex fluid field driving the particles in the azimuthal direction.

With the expression for the mobility tensor given in Appendix~\ref{app:mu-tensor}, its divergence reduces to a simple derivative with respect to the distance $z$ of the particles from the coverslip surface. The equations of motions \eqref{eq:langevin-laboratory} thus become
\begin{equation}
\frac{\dd \ri}{\dd t} = -\sum_{j=1}^N \vb*{\mu}_{ij} \vb*\nabla_j U_{\rm opt}(\vb*r_j,t) + 
\kb T \frac{\dd \murpbperp(z_i)}{\dd z_i} \vb*{e}_z + \vb*{\zeta}_i(t) \, ,
\label{eq:langevin-laboratory-2}
\end{equation}
In the experiments, we found that the particles are at a distance $\bar z=1.1\lambda$ from the coverslip surface with only small fluctuations in the $z$-direction. Here 
\begin{equation}
\lambda=\frac{2\pi R}{N_{\rm tr}}
\label{eq:lambda}
\end{equation}
is the distance between neighboring optical traps, that is the wavelength of the optical potential.
In the simulations, we
fixed $z_i=\bar z$ for all $i=1,\dots,N$, implying that the second term on the right hand side 
of Eq.~\eqref{eq:langevin-laboratory-2} does not influence the dynamics.

As we can neglect the motion in radial direction also, see the discussion after Eq.~\eqref{eq:Ust-1}, 
we project Eq.~\eqref{eq:langevin-laboratory-2} onto the
azimuthal direction, i.e.\ onto the axis given by the unit vector $\hat e_{\varphi_i}=(-\sin\varphi_i,\cos\varphi_i)$. 
We thus obtain the following equations of motions for the azimuthal angles,
\begin{equation}
R\frac{\dd\varphi_i}{\dd t} = -\hat e_{\varphi_i}\cdot\sum_{j=1}^N \vb*{\mu}_{ij} \vb*\nabla_j U_{\rm opt}(\vb*r_j,t)
+ \hat e_{\varphi_i}\cdot\vb*{\zeta}_i(t) \, ,
\label{eq:langevin-varphi}
\end{equation}
The total force given by the sum in this equation is evaluated from the particle positions
in the $x,y$-plane at $z=\bar z$.
Note that for weaker confining potentials in the radial direction, a HI induced particle
pairing was reported, where the two particles forming pairs align almost along the radial direction \cite{Sokolov/etal:2011}. 
However, in our experiments we did not observe any indication of such pairing, see also the videos referred to in 
Sec.~\ref{subsec:potential-shape} after Eq.~\eqref{eq:Ust-1}.

In the corotating frame, Eq.~\eqref{eq:langevin-varphi} becomes, with $\bm\omega=\omega\hat e_z$ [$\hat e_z$ is the unit vector in $z$-direction, see Fig.~\ref{fig:experimental_setup}(b)],
\begin{equation}
R\frac{\dd\varphi_i'}{\dd t} = -\omega R
-\hat e_{\varphi_i'}\cdot\sum_{j=1}^N \bm\mu_{ij}'\bm\nabla^{\,\prime}_j U_{\rm opt}'(\bm{r}^{\,\prime}_j,t)
+\hat e_{\varphi_i'}\cdot \bm\zeta_i(t)\,.
\label{eq:langevin-varphiprime}
\end{equation}

The hard-sphere interaction
implies that the distances $R\Delta\varphi$ between neighboring particles cannot be smaller than $2a$
and that the particles keep their order (single-file transport). 
These constraints are taken into account in the simulation by the procedure proposed by Scala {\it et al.} \cite{Scala/etal:2007}
and adopted to periodic potentials \cite{Ryabov/etal:2019}.

\begin{figure}[t!]
\centering
\includegraphics[scale=1]{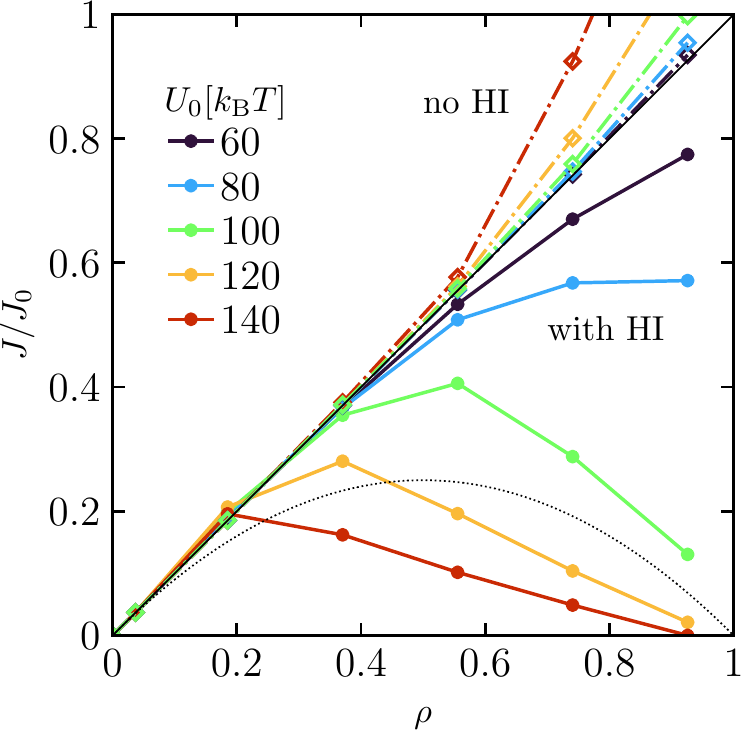}
\caption{Simulated fundamental diagrams for a ring of radius $R=20.2\,\si{\mu m}$ with
$N_{\rm tr}=27$ optical traps that are rotated with an angular frequency $\omega=0.63\,\si{rad/s}$. The potential amplitude modulation is $\xi=0.22$ 
[see Eqs.~\eqref{eq:Ureal_laboratory}, \eqref{eq:Ureal_corotating}], and results are shown for various mean barrier heights $U_0$.
For the simulations without HI,
the currents increase with $\rho$ and become larger than the current $J_0\rho$ of independent particles (straight solid line with slope one). When taking HI into account, the stationary current $J(\rho)$ becomes smaller than $J_0\rho$ for large $\rho$. For barrier heights $U_0\gtrsim 80$, it runs through a maximum. Strong decrease of the currents with $\rho$ is obtained for larger $U_0$ and reflect a jamming behavior.}
\label{fig:simresults_with_and_without_HI}
\end{figure}

\section{HI induced enhancement of barriers and implications for jamming}
\label{sec:current-density-relations}
Simulations for the system with radius $R=20.2\,\si{\mu m}$ were carried out with and without consideration of HI. 
Figure~\ref{fig:simresults_with_and_without_HI} shows
current-density relations obtained from these simulations
for various
barrier heights $U_0$ in a range investigated in the experiments. The density is given by the filling factor of the traps, 
\begin{equation}
\rho=\frac{N}{N_{\rm tr}}\,.
\end{equation}
The currents were obtained from particle trajectories in the many-body system 
by first determining azimuthal velocities over time for all particles (same in experiments).
After averaging, we obtain the mean velocity $\langle v\rangle$ in azimuthal direction 
and the current $J=N\langle v\rangle/(2\pi R)$. 
We normalized these currents with respect to
\begin{equation}
J_0=\frac{v_0N_{\rm tr}}{2\pi R}\,,
\end{equation}
which is the  single-particle
current $v_0/(2\pi R)$ times the number of traps;  $v_0$ is the mean single-particle velocity shown in Fig.~\ref{fig:v0}. 
The multiplication with $N_{\rm tr}$ implies that $J/J_0=\rho$ for independent particles. Accordingly, the simulated currents in 
Fig.~\ref{fig:simresults_with_and_without_HI}  approach the solid black line with slope one for small $\rho$.

It is important to note that the currents in Fig.~\ref{fig:simresults_with_and_without_HI} 
are the ones in the corotating frame, i.e.\ the currents were determined
based on Eq.~\eqref{eq:langevin-varphiprime}. In particular, this means that in the absence of the optical potential, the particles would rotate counterclockwise with an angular velocity $\omega R$. For an optical potential with very high barriers, where  the particles are carried along with the rotating traps, the current would be zero.

Comparing the results with and without HI shows that the presence of the HI is decisive for a suppression of the current at high particle densities to occur, 
as it is seen also in the experiments, see Fig.~\ref{fig:simresults_with_and_without_HI}. In the absence of HI, the currents increase with density at large $
\rho$, which can be attributed to a cluster speed-up effects discussed in Refs.~\cite{Antonov/etal:2022a, Antonov/etal:2022b, Lips/etal:2021}.

How can we understand that HIs lead to a slowing down of the particle transport and a jamming behavior for
large $U_0$ at high densities? To answer this question, we estimate how the HI modifies the motion for a particle
that is typically close to the minimum of an optical trap and that has two neighbors at a distance of about $\lambda$, i.e.\
close to the trap minima next to it. As the mobilities in Eq.~\eqref{eq:langevin-varphi} decrease with distance, we consider the sum of HI induced forces to be dominated by the contribution of these neighboring particles. Also, we neglect the relatively weak
flow effects caused by the coverslip surface, i.e.\ we use the mobility tensors $\bm\mu_{ij}$ at the Rotne-Prager level
with components
\begin{equation}
\frac{(\murp_{ij})_{\alpha\beta}}{\mu_0}=\left\{\begin{array}{cl}
\delta_{\alpha\beta}\,, & i=j\,,\\[1.5ex]
\hspace{-3em}\displaystyle\frac{3}{4} \frac{a}{r_{ij}} \left( \delta_{\alpha \beta} + \frac{r^\alpha_{ij} r^\beta_{ij}}{r^2_{ij}} \right)\\[3.5ex] 
\hspace{2em}\displaystyle+ \frac{1}{2}\frac{a^3}{r^3_{ij}} \left( \delta_{\alpha \beta} - 3 \frac{r^\alpha_{ij} r^\beta_{ij}}{r^2_{ij}} \right), 
& i \neq j \, .
\end{array}\right.
\label{eq:mu-rp-level}
\end{equation}
Here $r_{ij}^\alpha=(\bm r_j-\bm r_i)_\alpha$, and the non-diagonal term follows from Eq.~\eqref{eq:rp-tensor}
when taking into account
$\mu_0=1/(6\pi\eta a)$ according to Stokes friction law ($\eta=\rho_{{\rms H}_2{\rms O}}\nu$ is the dynamic viscosity).

\begin{figure}[t!]
\includegraphics[scale=1]{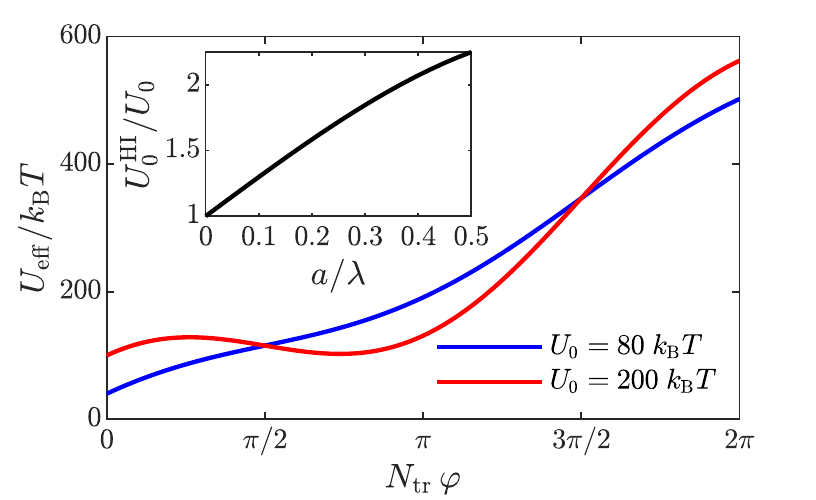}
\centering
\caption{Effective potential \eqref{eq:Ueff} in the corotating frame for two values of the barrier height $U_0$.
The data are shown for the same number of optical traps $N_{\rm tr}=27$, ring radius 
$R=20.2\,\si{\mu m}$, and angular frequency $\omega=0.63\,\si{rad/s}$ as in Fig.~\ref{fig:simresults_with_and_without_HI}, yielding a tilting torque 
$\omega R^2/N_{\rm tr}\mu_0\cong 73.5\kb T$. For this tilting torque, the effective potential starts to exhibit minima and maxima
when $U_0\gtrsim 147\,\kb T$. Accordingly, for $U_0=80\,\kb T$, the tilting is overcritical and $U_{\rm eff}$ varies
monotonically with $\varphi'$.  For $U_0=200\,\kb T$ by contrast, the tilting is undercritical and $U_{\rm eff}$ shows 
potential barriers. In the inset, we show the HI-induces barrier enhancement as predicted by Eq.~\eqref{eq:force-estimate},
i.e.\ $U_0^{\rm HI}/U_0\approx (1+3a/\lambda-2a^3/\lambda^3)$.}
\label{fig:tilting_transition}
\end{figure}

The particle and the two neighboring ones at distance $\lambda$ are located essentially along a line (for neighboring traps, the curvature of the ring can be neglected) and the optical forces exerted on them are almost in the direction of this line.
We thus approximate the sum in Eq.~\eqref{eq:langevin-varphi} by considering a one-dimensional geometry in a direction 
$\hat{\bm e}_\alpha$,
yielding 
\begin{subequations}
\begin{align}
&\hspace{-8ex}\sum_{j=1}^N \vb*{\mu}_{ij} \vb*\nabla_j U_{\rm opt}(\vb*r_j,t)\nonumber\\[-1ex]
&\approx
(\bm\mu_{ii})_{\alpha\alpha} \frac{\partial U_{\rm opt}}{\partial x_{i,\alpha}}
+(\bm\mu_{i,i-1})_{\alpha\alpha} \frac{\partial U_{\rm opt}}{\partial x_{i-1,\alpha}}\\
\nonumber &\hspace{3.5ex}+(\bm\mu_{i,i+1})_{\alpha\alpha} \frac{\partial U_{\rm opt}}{\partial x_{i+1,\alpha}}\nonumber\\
&\approx [\mu_0+2(\bm\mu_{i,i+1})_{\alpha\alpha}] \frac{\partial U_{\rm opt}}{\partial x_{i,\alpha}}\,.
\end{align}
\end{subequations}
When going from the second to the third line we used that the particles 
have mutual distances $\lambda$ and that
the optical potential is $\lambda$-periodic. As $\bm\mu_{ij}$ depends only on the distance between particles, we 
used also  $\bm\mu_{i,i-1}=\bm\mu_{i,i+1}$. Inserting the diagonal element of $(\bm\mu_{i,i+1})$ from 
Eq.~\eqref{eq:mu-rp-level}, we obtain
\begin{align}
\sum_{j=1}^N \vb*{\mu}_{ij} \vb*\nabla_j U_{\rm opt}(\vb*r_j,t)\approx
\mu_0 \left(1 + 3\frac{a}{\lambda} - 2\frac{a^3}{\lambda^3} \right)
\frac{\partial U_{\rm opt}}{\partial x_{i,\alpha}}\,.
\label{eq:force-estimate}
\end{align}

\begin{figure*}[t!]
\includegraphics[width=\textwidth]{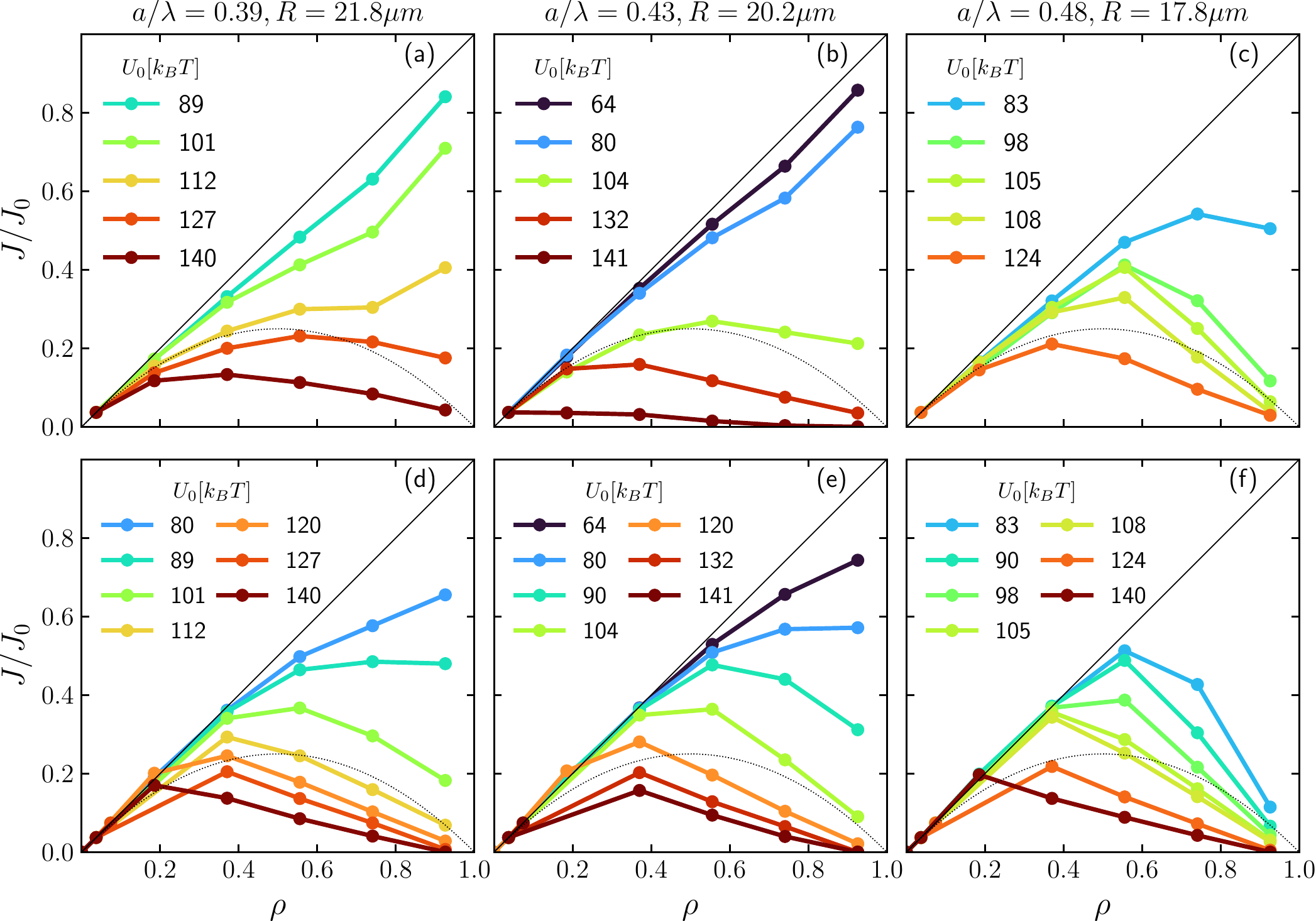}
\caption{Measured fundamental diagrams (upper row) in comparison with simulated ones (lower row) for three different ring radii $R$ (or particle radius $a$ in units of the wavelength $\lambda$). Results are shown for a fixed number $N_{\rm tr}=27$ of optical traps and
various mean barrier heights $U_0$, and the potential
amplitude modulation is $\xi=0.22$ in all graphs. The angular frequency $\omega$ for the rotation of the optical traps is adjusted such that $\omega R^2\cong257\,\si{{\mu m}^2/s}$ is constant, implying a constant tilting of the effective potential \eqref{eq:Ueff}, i.e.\ a constant drag torque in the corotating frame. The measured and simulated currents show qualitatively similar dependencies on $\rho$, with jamming characteristics appearing at large $\rho$ for sufficiently high mean barriers $U_0$. The value $U_0$ 
for the onset of the jamming behavior (transition from over- to undercritical tilting) shifts to smaller values for larger $a/\lambda$, 
as expected from Eq.~\eqref{eq:force-estimate}.}
\label{fig:exp_sim_results_j}
\end{figure*}

In the absence of HI, only the first term equal to the one in the bracket on the right hand side of Eq.~\eqref{eq:force-estimate} would 
be present. For our estimate based on the Rotne-Prager expression for the mobility tensor, the particle radius should be significantly smaller than $\lambda$. In this regime, the additional two terms in the bracket give an additional contribution larger than zero. 
For the radius $R=20.2\,\si{\mu m}$, we have $(a/\lambda)\cong0.43$, and the bracket gives a value of about two. 
This means that the optical potential has an effectively higher mean barrier height $U_0$.
It is intuitively clear that such barrier enhancement should slow down the motion but it is not immediately evident why it leads to the jamming behavior. 

The jamming like behavior can be understood from the effective mean external potential $U_{\rm eff}(\varphi')$
in the azimuthal direction in the corotating frame. In the absence of the driving, this potential is given
by $U_\varphi^{\rm st}(\varphi')$ from Eq.~\eqref{eq:potential-static} 
 with an HI-enhanced barrier height $U_0^{\rm HI}$ (and we set the irrelevant constant $c_0=0$). The flow-driving
given by $(-\bm\omega\times\bm r^{\,\prime})\simeq (-\omega R\varphi')\hat{\bm e}_{\varphi'}$ in 
Eq.~\eqref{eq:langevin-corotating} corresponds to a constant torque
$(-\omega R^2/\mu_0)$. Accordingly,
$U_\varphi^{\rm st}(\varphi')$ becomes tilted in $(-\varphi')$-direction, i.e.\
\begin{equation}
U_{\rm eff}(\varphi')
=\frac{U_0^{\rm HI}}{2}\cos(N_{\rm tr}\varphi')+\frac{\omega R^2}{\mu_0}\varphi'\,.
\label{eq:Ueff}
\end{equation}
For a given tilt, as in our experiment, this potential exhibits minima only for sufficiently large barrier heights, i.e.\ in the regime commonly
referred to as undercritical tilting. In that case, potential wells exist that, when occupied by a particle,
form an obstacle for the motion of nearby particles (blocking effect) and lead to jamming at high particle densities.

Figure~\ref{fig:tilting_transition} shows the transition from over- to undercritical tilting for the potential in Eq.~\eqref{eq:Ueff}.
For $U_0^{\rm HI}$ smaller than a critical $U_{0{\rm c}}^{\rm HI}$, the particles can slide down in the $(-\varphi)$-direction, while
for $U_0^{\rm HI}>U_{0{\rm c}}^{\rm HI}$ they have to surmount potential barriers in thermally activated rare events.
The critical barrier for passing from over- to undercritical tilting with increasing $U_0^{\rm HI}$ is
$U_{0{\rm c}}^{\rm HI}=2\omega R^2/N_{\rm tr}D_0\cong 147\,k_{\rm B}T$. 
Accordingly, due to the HI-enhanced barrier height, 
the regime of undercritical tilting can occur already for $U_0$ much smaller than the critical value,
and this leads to jamming at large $\rho$ for comparatively small barrier heights,
see Fig.~\ref{fig:simresults_with_and_without_HI}.

According to the estimate in Eq.~\eqref{eq:force-estimate}, the barrier enhancement should increase with the ratio
$a/\lambda$ for $a/\lambda\le1/2$, see the inset of Fig.~\ref{fig:tilting_transition}, where we 
show $U_0^{\rm HI}/U_0\approx (1 + 3a/\lambda - 2a^3/\lambda^3)$. 
As a consequence, we expect the transition from over- to undercritical tilting to shift towards smaller barrier heights $U_0$ with decreasing $\lambda$ or decreasing ring 
radius $R$ [see Eq.~\eqref{eq:lambda}]. 

To test this, we have carried out other experiments and corresponding simulations for two other radii $R=17.8\,\si{\mu m}$ and $R=21.8\,\si{\mu m}$. To keep the tilting in Eq.~\eqref{eq:Ueff} and hence the drag torque on the particles constant, we 
adjusted $\omega$ for each radius such that $\omega R^2$ remains constant.
The results for the current-density relations (fundamental diagrams) for all three studied radii
are shown in Fig.~\ref{fig:exp_sim_results_j}. In the upper row, the experimental results are displayed. 
For all three radii, a jamming behavior is seen at large $\rho$ and sufficiently strong barrier height $U_0$. When comparing curves for similar barrier heights, one notices that the jamming becomes more pronounced for smaller radii $R$ or larger 
ratios $a/\lambda$. The corresponding simulated results in Fig.~\ref{fig:exp_sim_results_j}
show an analogous behavior. 

In particular, the onset of the jamming behavior is shifted in agreement with our expectation: For $R=17.8\,\si{\mu m}$ ($a/\lambda=0.48$) it occurs at a barrier height $U_0$ of about $80\,\kb T$, while for $R=21.8\,\si{\mu m}$ ($a/\lambda=0.39$) it is at a higher
$U_0\simeq 120\,\kb T$. The shift  given by the term in parentheses on the right-hand side of
 Eq.~\eqref{eq:force-estimate} would be smaller than this, but we should not expect to obtain a quantitative agreement from our simple estimate. 

When comparing experimental and simulated data for the same $U_0$ values, we see also that there is no quantitative match. This can be due to the simplifications in our modeling, where we have considered the particle motion to be perfectly confined along the azimuthal direction, and where we treated the HIs approximately by the Blake tensor at
the Rotne-Prager level.

\section{Flow-driven versus force-driven systems}
\label{sec:flow-vs-force-driving}
\begin{figure}[t!]
\centering
\includegraphics[scale=1]{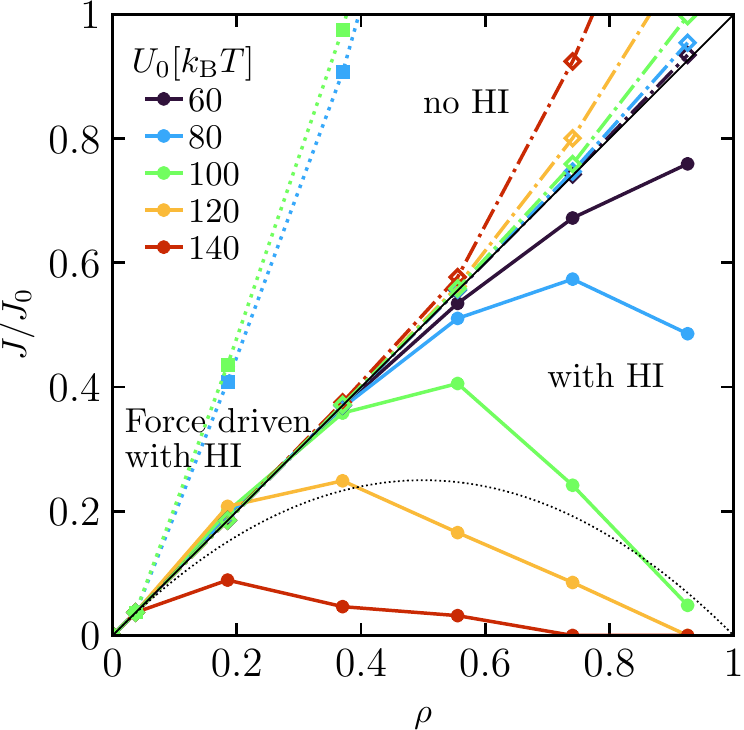}
\caption{Simulated fundamental diagrams for the same parameters as in Fig.~\ref{fig:simresults_with_and_without_HI}, 
but without modulation of the potential amplitude ($\xi=0$). The results with and without HI under flow-driving show 
essentially the same behavior as in Fig.~\ref{fig:simresults_with_and_without_HI}. This implies that the 
jamming behavior is not caused by the amplitude modulation but by the HI. For a force-driven system with HI, 
the currents are much larger than in the flow-driven system without HI.}
\label{fig:j_simulations_noAmodulation}
\end{figure}

As mentioned in the introduction, previous findings reported on a barrier reduction effect induced by HIs. These findings
were obtained for force-driven systems. For our setup, we have performed simulations also when the particles are driven by a force instead of a flow. In these simulations, we have also switched off the amplitude modulation, which is not a crucial factor for our results reported above. The fundamental diagrams obtained from these simulations are shown in Fig.~\ref{fig:j_simulations_noAmodulation}. In the flow-driven system, the jamming behavior is seen in the presence of HI. In
the absence of HI, we obtain even a current enhancement compared to $J_0\rho$ at larger particle densities. 
This can be explained by the fact that the effective potential in Eq.~\eqref{eq:Ueff}
has steep slopes (large driving forces) in certain regions. When increasing the density $\rho$, these regions with large driving
forces become more strongly populated, leading to the enhancement of the current. 

In the force-driven system by contrast, the results in Fig.~\ref{fig:j_simulations_noAmodulation}
do not reflect a jamming behavior in the presence of HI and the currents are much larger. Why are
the currents so much different compared to a flow-driven system?

In the force-driven case, we have a drag force $\bm f_i=\bm f^{\rm drag}(\bm r_i)=-\mu_0^{-1}\bm\omega\times\bm r_i$ acting on particle $i$.
The equations of motion \eqref{eq:langevin-corotating} then become
\begin{equation}
\frac{\dd {\bm{r}}_i}{\dd t}=\sum_{j=1}^{N}
\Big\{k_{\rm B}T\bm\nabla_j \bm\mu_{ij}
+\bm\mu_{ij}[-\bm\nabla_j U_{\rm opt}(\bm{r}_j,t)+\bm f_j]\Bigr\}
+ \bm\zeta_i\,.
\label{eq:langevin-corotating-force-driving}
\end{equation}
For the comparison with the flow-driven system, we can ignore the experimental way of generating the flow-driving, i.e.\ disregard the prime on the coordinates in Eq.~\eqref{eq:langevin-corotating}. The effect of the mobility tensors ${\bm\mu}_{ij}$ on these drag forces would then be the same as on the forces $(-\bm\nabla_j U_{\rm opt})$ of the optical potential, implying that both the barrier height and the drag force amplitude are enhanced by a factor as estimated in Eq.~\eqref{eq:force-estimate}. This means that no transition between the regimes of over- and undercritical is occuring with increasing $U_0$. Instead, the system stays in the regime of overcritical tilting and the density dependence of the 
current is almost unaffected when changing $U_0$.

The much higher currents in the force-driven system compared to the flow-driven one can be understood from the fact that the mean 
barrier height and drag force amplitude are enhanced by HI in the same manner. When neglecting the divergences of the mobility 
tensors and  when taking into account that the experimental conditions refer to a situation of low noise ($k_{\rm B}T\ll U_0$), the 
enhancement amounts to a mere speeding-up of all velocities $\dot{\bm r}_i$ by the same factor. This is not the case in the flow-
driven system, where the amplitudes of the optical forces $(-\bm\nabla_i U_{\rm opt})$ are HI-enhanced but not that of the 
flow-driving forces $\propto\bm\omega\times\bm r_i$.

\section{Summary and conclusions}
\label{sec:conclusions}
Transport over potential barriers is omnipresent in nature. Here we have reported experimental and theoretical results on the question how
HIs influence such transport in the case of Brownian motion of hard spheres in a periodic potential. 

In our experiments we designed a setup of rotating optical traps that allowed us to study the particle transport under a vortex flow-field. For deep optical traps, we found that the currents strongly decrease with the particle density, reflecting a jamming behavior. Stokesian dynamics simulations of the setup showed that this behavior is caused by HIs. We interpreted this finding by an effective HI-induced enhancement of the barriers between the optical traps, which leads to a transition from an overcritical to an undercritical tilting of the optical potential. The enhancement  is contrary to what has been reported for the impact of HI on potential barriers in force-driven systems. Comparison of flow- to force-driving in our setup also shows striking differences. They are caused by an effective enhancement of the force amplitude going along with the barrier increase. As a consequence, no transition of the tilting characteristics is induced by the HI and the currents are only weakly affected by the depths of the optical traps.

Our interpretation for the jamming as a consequence of a HI-induced transition from over- to undercritical tilting of the optical 
potential provides an understanding of the observed behavior. The formula \eqref{eq:force-estimate} derived for estimating the HI-
induced barrier enhancement, however, should not be considered as giving quantitative predictions.
To this end, the theoretical treatment needs to be improved. 

One possible improvement concerns the consideration of fluctuations in 
particle positions and of particles at distances beyond neighboring potential wells.
Another concerns the Rotne-Prager level of 
description, even when taking into account the additional terms in the Blake tensor to account for the flow effects from the coverslip 
surface. Not considered yet are in particular effects when particles come close to each other, which
would require to take into account terms of higher order in the multipole expansion of mobilities and lubrication effects associated with
a rotational motion of the particles. This can be done based on the mobility method by using the expressions derived 
in Ref.~\cite{Durlofsky/etal:1987} for lubrication forces between the particles, and in
Ref.~\cite{Swan/Brady:2007} to take into account lubrication effects of the particles with the coverslip.
Another possible approach is to resort to
large-scale simulation methods with explicit modeling of fluid flows in multiparticle collision dynamics \cite{Howard/etal:2019}.

From the experimental side, the explored system
can be extended in various ways. For example, the present work is centered on the flow-driven properties of monodisperse 
systems. One could ask how the collective dynamics and the reported jamming effect are altered in polydisperse systems
with particles of different sizes.
The simplest case would be to include one or few differently sized particles
within the ring  and investigate 
how these inclusions behave by varying their relative fraction and the depth of the potential wells. 
Smaller or larger particles will see a different potential well than their neighbors, 
and this will affect their jumping over the potential barriers.   

Further, modern colloid science allows to prepare monodisperse particles with a shape that departs from the spherical one, such as ellipsoids \cite{Champion2007} or even cubes \cite{Rossi2010}. Using anisotropic colloids will increase the complexity of the system when considering the generated HIs but, on the other hand, will allow one to 
extract experimentally also the relative orientation of the particles within the ring. This further piece of information could be used to determine whether, in the steady state, such particles synchronize their rotational motion due to HIs under the driving. 

The nature of the particles can also be changed, for example by using monodisperse paramagnetic colloids, which could be easily 
manipulated via magnetic fields in this geometry \cite{Ortiz2018, Gerloff2020}. One could increase the density of particles to induce 
jamming, but use a rotating field \cite{Tierno2007, Martinez2015} with opposite chirality to apply 
a torque which would un-jam the system.

Finally, instead of changing the particles, the AOD can be programmed in such a way to introduce a controlled degree of spatial or 
temporal disorder  in the potential landscape, for example by reducing or increasing the time that the laser beam visits one trap. 
One can also design more complex optical paths than the circular one, like elliptical or square where the presence of
sharp corners could induce earlier the jamming behavior or could act as a sink of particles to be later released at high angular speed. 

Exploring these interesting options will foster our understanding of HI effects in natural systems and could open 
ways towards their exploitation in technological applications. 

\section*{Acknowledgements}
We sincerely thank G.\ N\"agele and H.\ Stark for advice on the treatment of hydrodynamic interactions. This project has received funding from the European Research Council (ERC) under the European Union's Horizon 2020 research and innovation program (grant agreement no.\ 811234). P.T. acknowledges support from the Generalitat de Catalunya under Program ``ICREA Acad\`emia''. A.R. and P.M. gratefully acknowledge financial support by the Czech Science Foundation (Project No.\ 20-24748J) and the Deutsche Forschungsgemeinschaft (Project Nos.\ 397157593, 355031190). We sincerely thank the members of the DFG Research Unit FOR 2692 for fruitful discussions.

\appendix

\section*{APPENDIX}

\section{Blake tensor at the\\ Rotne-Prager level \cite{Hansen/etal:2011}}
\renewcommand{\theequation}{A\arabic{equation}}\setcounter{equation}{0}
\label{app:mu-tensor}
The Blake tensor at the Rotne-Prager level decomposes into a self and an interaction part,
\begin{equation}
\vb*{\mu}_{ij} = \muself (z_i) \delta_{ij} + (1-\delta_{ij})\murpb(\ri, \rj)\,.
\end{equation}
The self part depends only on the distance $z$ of a particle from the coverslip surface,
\begin{equation}
\muself (z) = \begin{pmatrix}
\mu_{\parallel}^{\rms RPB}(z) & 0 & 0 \\
0 & \mu_{\parallel}^{\rms RPB}(z) & 0 \\
0 & 0 & \mu_{\perp}^{\rms RPB}(z) \\
\end{pmatrix}
\end{equation}
\begin{subequations}
\begin{equation}
\mu_{\parallel}^{\rms RPB}(z) = \mu_0 \left(1 - \frac{9a}{16z} + \frac{1}{8}\left(\frac{a}{z}\right)^3 \right)
\end{equation}
\begin{equation}
\mu_{\perp}^{\rms RPB}(z) = \mu_0 \left(1 - \frac{9a}{8z} + \frac{1}{2}\left(\frac{a}{z}\right)^3 \right)\,.
\end{equation}
\end{subequations}
The interaction part is given by the ``curvature correction'' (second order term of multipole expansion)
of the Blake tensor,
\begin{equation}
\murpb (\ri, \rj) = \left( 1 + \frac{a^2}{6} \nabla_{\ri}^2 + \frac{a^2}{6} \nabla_{\rj}^2 \right)\mub(\ri, \rj)\,.
\end{equation}
The Blake tensor is
\begin{align}
\mub (\ri, \rj) = &\vb*{\mu}^{\rms O}(\ri - \rj) - \vb*{\mu}^{\rms O}(\ri - \rjb)\nonumber\\
&+ \vb*{\mu}^{\rms D}(\ri - \rjb) - \vb*{\mu}^{\rms SD}(\ri - \rjb)\,,
\end{align}
where  $\rjb = (x_j, y_j, -z_j)$, and $\bm\mu^{\rms O}$ is the Oseen tensor, $\bm\mu^{\rms D}$ the Stokes doublet, and $\bm\mu^{\rms SD}$ the
 source doublet. These tensors are given by
\begin{subequations}
\begin{equation}
\vb*{\mu}_{\alpha \beta}^{\rms O}(\vb*{r}) = \frac{1}{8 \pi \eta r} \left(\delta_{\alpha\beta} + \frac{r_\alpha r_\beta}{r^2} \right)
\end{equation}
\begin{equation}
\vb*{\mu}^{\rms D}_{\alpha\beta}(\vb*{s}) = \frac{2 z_j^2(1-2\delta_{\beta z})}{8 \pi \eta}
\left( \frac{\delta_{\alpha\beta}}{s^3} - \frac{3 s_\alpha s_\beta}{s^5} \right)
\end{equation}
\begin{align}
\vb*{\mu}^{\rms SD}_{\alpha\beta}(\vb*{s})=&\frac{2 z_j(1-2\delta_{\beta z})}{8 \pi \eta}\\[1ex]
&{}\times\left( \frac{\delta_{\alpha\beta}s_z}{s^3} - \frac{\delta_{\alpha z}s_\beta}{s^3} 
+ \frac{\delta_{\beta z}s_\alpha}{s^3} - \frac{3 s_\alpha s_\beta s_z}{s^5} \right)\,,
\nonumber
\end{align}
\end{subequations}
where  $\vb*{s}=\ri-\rjb$, $r = \vert \vb*{r} \vert$, and $\eta$ is the dynamic viscosity
($\eta=\rho_{{\rms H}_2{\rms O}}\nu$ in the experiment with $\nu$ the kinematic viscosity of water).

\pagebreak
The explicit expression for $\murpb$ is
\begin{equation}
\murpb(\ri, \rj) = \murp(\ri - \rj) - \murp(\ri-\rjb) + \vb*{\Delta\mu}(\ri, \rj) \, ,
\end{equation}
where
\begin{equation}
\murp(\vb*{r}) = \frac{1}{8 \pi \eta r} \left[ \vb*{1} + \frac{\vb*{r} \otimes \vb*{r}}{r^2} \right]
+ \frac{a^2}{4 \pi \eta r^3} \left[ \frac{\vb*{1}}{3} - \frac{\vb*{r} \otimes \vb*{r}}{r^2} \right] \,,
\label{eq:rp-tensor}
\end{equation}
\begin{subequations}
and  $\vb*{\Delta\mu}(\ri, \rj)$ has diagonal components ($\alpha=x,y$)
\begin{equation}
\vb*{\Delta \mu}_{\alpha\alpha} = \frac{1}{4 \pi \eta} \left[ 
\frac{-z_i z_j}{s^3} \left( 1 - 3\frac{s_\alpha^2}{s^2} \right)
+ \frac{a^2 s_z^2}{s^5} \left( 1 - 5 \frac{s_\alpha^2}{s^2} \right)
\right] \, ,
\end{equation} 
\begin{equation}
\vb*{\Delta \mu}_{zz} = \frac{1}{4 \pi \eta} \left[ 
\frac{z_i z_j}{s^3} \left( 1 - 3\frac{s_z^2}{s^2} \right)
- \frac{a^2 s_z^2}{s^5} \left( 3 - 5 \frac{s_z^2}{s^2} \right)
\right] \,,
\end{equation}
and off-diagonal elements ($\alpha=x,y$, $\beta=x,y$)
\begin{equation}
\vb*{\Delta \mu}_{\alpha\beta} = \frac{1}{4 \pi \eta} \left[ 
\frac{3 z_i z_j s_\alpha s_\beta}{s^5} 
- 5 a^2 \frac{s_\alpha s_\beta s_z^2}{s^7}
\right] \,,
\end{equation}
\begin{equation}
\vb*{\Delta \mu}_{\alpha z} = \frac{1}{4 \pi \eta} \left[ 
\frac{z_j s_\alpha}{s^3} \left( 1 - 3\frac{z_i s_z}{s^2} \right)
- \frac{a^2 s_\alpha s_z}{s^5} \left( 2 - 5 \frac{s_z^2}{s^2} \right)
\right] \,,
\end{equation}
\begin{equation}
\vb*{\Delta \mu}_{z \alpha} = \frac{1}{4 \pi \eta} \left[ 
\frac{z_j s_\alpha}{s^3} \left( 1 + 3\frac{z_i s_z}{s^2} \right)
- 5\frac{a^2 s_\alpha s_z^3}{s^7}
\right] \,.
\end{equation}
\end{subequations}


\begin{thebibliography}{77}%
\makeatletter
\providecommand \@ifxundefined [1]{%
 \@ifx{#1\undefined}
}%
\providecommand \@ifnum [1]{%
 \ifnum #1\expandafter \@firstoftwo
 \else \expandafter \@secondoftwo
 \fi
}%
\providecommand \@ifx [1]{%
 \ifx #1\expandafter \@firstoftwo
 \else \expandafter \@secondoftwo
 \fi
}%
\providecommand \natexlab [1]{#1}%
\providecommand \enquote  [1]{``#1''}%
\providecommand \bibnamefont  [1]{#1}%
\providecommand \bibfnamefont [1]{#1}%
\providecommand \citenamefont [1]{#1}%
\providecommand \href@noop [0]{\@secondoftwo}%
\providecommand \href [0]{\begingroup \@sanitize@url \@href}%
\providecommand \@href[1]{\@@startlink{#1}\@@href}%
\providecommand \@@href[1]{\endgroup#1\@@endlink}%
\providecommand \@sanitize@url [0]{\catcode `\\12\catcode `\$12\catcode
  `\&12\catcode `\#12\catcode `\^12\catcode `\_12\catcode `\%12\relax}%
\providecommand \@@startlink[1]{}%
\providecommand \@@endlink[0]{}%
\providecommand \url  [0]{\begingroup\@sanitize@url \@url }%
\providecommand \@url [1]{\endgroup\@href {#1}{\urlprefix }}%
\providecommand \urlprefix  [0]{URL }%
\providecommand \Eprint [0]{\href }%
\providecommand \doibase [0]{https://doi.org/}%
\providecommand \selectlanguage [0]{\@gobble}%
\providecommand \bibinfo  [0]{\@secondoftwo}%
\providecommand \bibfield  [0]{\@secondoftwo}%
\providecommand \translation [1]{[#1]}%
\providecommand \BibitemOpen [0]{}%
\providecommand \bibitemStop [0]{}%
\providecommand \bibitemNoStop [0]{.\EOS\space}%
\providecommand \EOS [0]{\spacefactor3000\relax}%
\providecommand \BibitemShut  [1]{\csname bibitem#1\endcsname}%
\let\auto@bib@innerbib\@empty
\bibitem [{\citenamefont {Happel}\ and\ \citenamefont
  {Brenner}(1973)}]{Happel/Brenner:1973}%
  \BibitemOpen
  \bibfield  {author} {\bibinfo {author} {\bibfnamefont {J.}~\bibnamefont
  {Happel}}\ and\ \bibinfo {author} {\bibfnamefont {H.}~\bibnamefont
  {Brenner}},\ }\href@noop {} {\emph {\bibinfo {title} {Low Reynolds Number
  Hydrodynamics}}}\ (\bibinfo  {publisher} {Noordhoff, Leiden},\ \bibinfo
  {address} {Noordhoff},\ \bibinfo {year} {1973})\BibitemShut {NoStop}%
\bibitem [{\citenamefont {Dhont}(1996)}]{Dhont:1996}%
  \BibitemOpen
  \bibfield  {author} {\bibinfo {author} {\bibfnamefont {J.~K.~G.}\
  \bibnamefont {Dhont}},\ }\href@noop {} {\emph {\bibinfo {title} {An
  Introduction to Dynamics of Colloids}}}\ (\bibinfo  {publisher} {Elsevier,
  Amster- dam},\ \bibinfo {address} {Boston},\ \bibinfo {year}
  {1996})\BibitemShut {NoStop}%
\bibitem [{\citenamefont {Hess}\ and\ \citenamefont
  {Klein}(2006)}]{Hess/Klein:2006}%
  \BibitemOpen
  \bibfield  {author} {\bibinfo {author} {\bibfnamefont {W.}~\bibnamefont
  {Hess}}\ and\ \bibinfo {author} {\bibfnamefont {R.}~\bibnamefont {Klein}},\
  }\bibfield  {title} {\bibinfo {title} {Generalized hydrodynamics of systems
  of {Brownian} particles},\ }\href {https://doi.org/10.1080/00018738300101551}
  {\bibfield  {journal} {\bibinfo  {journal} {Rep. Prog. Phys.}\ }\textbf
  {\bibinfo {volume} {71}},\ \bibinfo {pages} {173} (\bibinfo {year}
  {2006})}\BibitemShut {NoStop}%
\bibitem [{\citenamefont {Lauga}\ and\ \citenamefont
  {Powers}(2009)}]{Lauga/Powers:2009}%
  \BibitemOpen
  \bibfield  {author} {\bibinfo {author} {\bibfnamefont {E.}~\bibnamefont
  {Lauga}}\ and\ \bibinfo {author} {\bibfnamefont {T.~R.}\ \bibnamefont
  {Powers}},\ }\bibfield  {title} {\bibinfo {title} {The hydrodynamics of
  swimming microorganisms},\ }\href
  {https://doi.org/10.1088/0034-4885/72/9/096601} {\bibfield  {journal}
  {\bibinfo  {journal} {Rep. Prog. Phys.}\ }\textbf {\bibinfo {volume} {72}},\
  \bibinfo {pages} {096601} (\bibinfo {year} {2009})}\BibitemShut {NoStop}%
\bibitem [{\citenamefont {Riedel}\ \emph {et~al.}(2005)\citenamefont {Riedel},
  \citenamefont {Kruse},\ and\ \citenamefont {Howard}}]{Riedel/etal:2005}%
  \BibitemOpen
  \bibfield  {author} {\bibinfo {author} {\bibfnamefont {I.~H.}\ \bibnamefont
  {Riedel}}, \bibinfo {author} {\bibfnamefont {K.}~\bibnamefont {Kruse}},\ and\
  \bibinfo {author} {\bibfnamefont {J.}~\bibnamefont {Howard}},\ }\bibfield
  {title} {\bibinfo {title} {A self-organized vortex array of hydrodynamically
  entrained sperm cells.},\ }\href {https://doi.org/10.1126/science.1110329}
  {\bibfield  {journal} {\bibinfo  {journal} {Science}\ }\textbf {\bibinfo
  {volume} {309}},\ \bibinfo {pages} {300} (\bibinfo {year}
  {2005})}\BibitemShut {NoStop}%
\bibitem [{\citenamefont {Pierce}\ \emph {et~al.}(2018)\citenamefont {Pierce},
  \citenamefont {Wijesinghe}, \citenamefont {Mumper}, \citenamefont {Lower},
  \citenamefont {Lower},\ and\ \citenamefont {Sooryakumar}}]{Pierce/etal:2018}%
  \BibitemOpen
  \bibfield  {author} {\bibinfo {author} {\bibfnamefont {C.~J.}\ \bibnamefont
  {Pierce}}, \bibinfo {author} {\bibfnamefont {H.}~\bibnamefont {Wijesinghe}},
  \bibinfo {author} {\bibfnamefont {E.}~\bibnamefont {Mumper}}, \bibinfo
  {author} {\bibfnamefont {B.~H.}\ \bibnamefont {Lower}}, \bibinfo {author}
  {\bibfnamefont {S.~K.}\ \bibnamefont {Lower}},\ and\ \bibinfo {author}
  {\bibfnamefont {R.}~\bibnamefont {Sooryakumar}},\ }\bibfield  {title}
  {\bibinfo {title} {Hydrodynamic interactions, hidden order, and emergent
  collective behavior in an active bacterial suspension},\ }\href
  {https://doi.org/10.1103/PhysRevLett.121.188001} {\bibfield  {journal}
  {\bibinfo  {journal} {Phys. Rev. Lett.}\ }\textbf {\bibinfo {volume} {121}},\
  \bibinfo {pages} {188001} (\bibinfo {year} {2018})}\BibitemShut {NoStop}%
\bibitem [{\citenamefont {Lauga}\ \emph {et~al.}(2006)\citenamefont {Lauga},
  \citenamefont {DiLuzio}, \citenamefont {Whitesides},\ and\ \citenamefont
  {Stone}}]{Lauga/etal:2006}%
  \BibitemOpen
  \bibfield  {author} {\bibinfo {author} {\bibfnamefont {E.}~\bibnamefont
  {Lauga}}, \bibinfo {author} {\bibfnamefont {W.~R.}\ \bibnamefont {DiLuzio}},
  \bibinfo {author} {\bibfnamefont {G.~M.}\ \bibnamefont {Whitesides}},\ and\
  \bibinfo {author} {\bibfnamefont {H.~A.}\ \bibnamefont {Stone}},\ }\bibfield
  {title} {\bibinfo {title} {Swimming in circles: Motion of bacteria near solid
  boundaries.},\ }\href {https://doi.org/10.1529/biophysj.105.069401}
  {\bibfield  {journal} {\bibinfo  {journal} {Biophys. J.}\ }\textbf {\bibinfo
  {volume} {90}},\ \bibinfo {pages} {400} (\bibinfo {year} {2006})}\BibitemShut
  {NoStop}%
\bibitem [{\citenamefont {Berke}\ \emph {et~al.}(2008)\citenamefont {Berke},
  \citenamefont {Turner}, \citenamefont {Berg},\ and\ \citenamefont
  {Lauga}}]{Berke/etal:2008}%
  \BibitemOpen
  \bibfield  {author} {\bibinfo {author} {\bibfnamefont {A.~P.}\ \bibnamefont
  {Berke}}, \bibinfo {author} {\bibfnamefont {L.}~\bibnamefont {Turner}},
  \bibinfo {author} {\bibfnamefont {H.~C.}\ \bibnamefont {Berg}},\ and\
  \bibinfo {author} {\bibfnamefont {E.}~\bibnamefont {Lauga}},\ }\bibfield
  {title} {\bibinfo {title} {Hydrodynamic attraction of swimming microorganisms
  by surfaces},\ }\href {https://doi.org/10.1103/PhysRevLett.101.038102}
  {\bibfield  {journal} {\bibinfo  {journal} {Phys. Rev. Lett.}\ }\textbf
  {\bibinfo {volume} {101}},\ \bibinfo {pages} {038102} (\bibinfo {year}
  {2008})}\BibitemShut {NoStop}%
\bibitem [{\citenamefont {Di~Leonardo}\ \emph {et~al.}(2011)\citenamefont
  {Di~Leonardo}, \citenamefont {Dell'Arciprete}, \citenamefont {Angelani},\
  and\ \citenamefont {Iebba}}]{Leonardo/etal:2011}%
  \BibitemOpen
  \bibfield  {author} {\bibinfo {author} {\bibfnamefont {R.}~\bibnamefont
  {Di~Leonardo}}, \bibinfo {author} {\bibfnamefont {D.}~\bibnamefont
  {Dell'Arciprete}}, \bibinfo {author} {\bibfnamefont {L.}~\bibnamefont
  {Angelani}},\ and\ \bibinfo {author} {\bibfnamefont {V.}~\bibnamefont
  {Iebba}},\ }\bibfield  {title} {\bibinfo {title} {Swimming with an image},\
  }\href {https://doi.org/10.1103/PhysRevLett.106.038101} {\bibfield  {journal}
  {\bibinfo  {journal} {Phys. Rev. Lett.}\ }\textbf {\bibinfo {volume} {106}},\
  \bibinfo {pages} {038101} (\bibinfo {year} {2011})}\BibitemShut {NoStop}%
\bibitem [{\citenamefont {Reichert}\ and\ \citenamefont
  {Stark}(2005)}]{Reichert/Stark:2005}%
  \BibitemOpen
  \bibfield  {author} {\bibinfo {author} {\bibfnamefont {M.}~\bibnamefont
  {Reichert}}\ and\ \bibinfo {author} {\bibfnamefont {H.}~\bibnamefont
  {Stark}},\ }\bibfield  {title} {\bibinfo {title} {Synchronization of rotating
  helices by hydrodynamic interactions},\ }\href
  {https://doi.org/10.1140/epje/i2004-10152-7} {\bibfield  {journal} {\bibinfo
  {journal} {Eur. Phys. J. E}\ }\textbf {\bibinfo {volume} {17}},\ \bibinfo
  {pages} {493} (\bibinfo {year} {2005})}\BibitemShut {NoStop}%
\bibitem [{\citenamefont {Vilfan}\ and\ \citenamefont
  {J\"ulicher}(2006)}]{Vilfan/Juelicher:2006}%
  \BibitemOpen
  \bibfield  {author} {\bibinfo {author} {\bibfnamefont {A.}~\bibnamefont
  {Vilfan}}\ and\ \bibinfo {author} {\bibfnamefont {F.}~\bibnamefont
  {J\"ulicher}},\ }\bibfield  {title} {\bibinfo {title} {Hydrodynamic flow
  patterns and synchronization of beating cilia},\ }\href
  {https://doi.org/10.1103/PhysRevLett.96.058102} {\bibfield  {journal}
  {\bibinfo  {journal} {Phys. Rev. Lett.}\ }\textbf {\bibinfo {volume} {96}},\
  \bibinfo {pages} {058102} (\bibinfo {year} {2006})}\BibitemShut {NoStop}%
\bibitem [{\citenamefont {Drescher}\ \emph {et~al.}(2009)\citenamefont
  {Drescher}, \citenamefont {Leptos}, \citenamefont {Tuval}, \citenamefont
  {Ishikawa}, \citenamefont {Pedley},\ and\ \citenamefont
  {Goldstein}}]{Drescher/etal:2009}%
  \BibitemOpen
  \bibfield  {author} {\bibinfo {author} {\bibfnamefont {K.}~\bibnamefont
  {Drescher}}, \bibinfo {author} {\bibfnamefont {K.~C.}\ \bibnamefont
  {Leptos}}, \bibinfo {author} {\bibfnamefont {I.}~\bibnamefont {Tuval}},
  \bibinfo {author} {\bibfnamefont {T.}~\bibnamefont {Ishikawa}}, \bibinfo
  {author} {\bibfnamefont {T.~J.}\ \bibnamefont {Pedley}},\ and\ \bibinfo
  {author} {\bibfnamefont {R.~E.}\ \bibnamefont {Goldstein}},\ }\bibfield
  {title} {\bibinfo {title} {Dancing volvox: Hydrodynamic bound states of
  swimming algae},\ }\href {https://doi.org/10.1103/PhysRevLett.102.168101}
  {\bibfield  {journal} {\bibinfo  {journal} {Phys. Rev. Lett.}\ }\textbf
  {\bibinfo {volume} {102}},\ \bibinfo {pages} {168101} (\bibinfo {year}
  {2009})}\BibitemShut {NoStop}%
\bibitem [{\citenamefont {Uchida}\ and\ \citenamefont
  {Golestanian}(2011)}]{Uchida/Golestanian:2011}%
  \BibitemOpen
  \bibfield  {author} {\bibinfo {author} {\bibfnamefont {N.}~\bibnamefont
  {Uchida}}\ and\ \bibinfo {author} {\bibfnamefont {R.}~\bibnamefont
  {Golestanian}},\ }\bibfield  {title} {\bibinfo {title} {Generic conditions
  for hydrodynamic synchronization},\ }\href
  {https://doi.org/10.1103/PhysRevLett.106.058104} {\bibfield  {journal}
  {\bibinfo  {journal} {Phys. Rev. Lett.}\ }\textbf {\bibinfo {volume} {106}},\
  \bibinfo {pages} {058104} (\bibinfo {year} {2011})}\BibitemShut {NoStop}%
\bibitem [{\citenamefont {Brumley}\ \emph {et~al.}(2012)\citenamefont
  {Brumley}, \citenamefont {Polin}, \citenamefont {Pedley},\ and\ \citenamefont
  {Goldstein}}]{Brumley/etal:2012}%
  \BibitemOpen
  \bibfield  {author} {\bibinfo {author} {\bibfnamefont {D.~R.}\ \bibnamefont
  {Brumley}}, \bibinfo {author} {\bibfnamefont {M.}~\bibnamefont {Polin}},
  \bibinfo {author} {\bibfnamefont {T.~J.}\ \bibnamefont {Pedley}},\ and\
  \bibinfo {author} {\bibfnamefont {R.~E.}\ \bibnamefont {Goldstein}},\
  }\bibfield  {title} {\bibinfo {title} {Hydrodynamic synchronization and
  metachronal waves on the surface of the colonial alga volvox carteri},\
  }\href {https://doi.org/10.1103/PhysRevLett.109.268102} {\bibfield  {journal}
  {\bibinfo  {journal} {Phys. Rev. Lett.}\ }\textbf {\bibinfo {volume} {109}},\
  \bibinfo {pages} {268102} (\bibinfo {year} {2012})}\BibitemShut {NoStop}%
\bibitem [{\citenamefont {Chakrabarti}\ and\ \citenamefont
  {Saintillan}(2019)}]{Chakrabarti/Saintillan:2019}%
  \BibitemOpen
  \bibfield  {author} {\bibinfo {author} {\bibfnamefont {B.}~\bibnamefont
  {Chakrabarti}}\ and\ \bibinfo {author} {\bibfnamefont {D.}~\bibnamefont
  {Saintillan}},\ }\bibfield  {title} {\bibinfo {title} {Hydrodynamic
  synchronization of spontaneously beating filaments},\ }\href
  {https://doi.org/10.1103/PhysRevLett.123.208101} {\bibfield  {journal}
  {\bibinfo  {journal} {Phys. Rev. Lett.}\ }\textbf {\bibinfo {volume} {123}},\
  \bibinfo {pages} {208101} (\bibinfo {year} {2019})}\BibitemShut {NoStop}%
\bibitem [{\citenamefont {Crocker}\ and\ \citenamefont
  {Grier}(1996)}]{Crocker1996}%
  \BibitemOpen
  \bibfield  {author} {\bibinfo {author} {\bibfnamefont {C.}~\bibnamefont
  {Crocker}}\ and\ \bibinfo {author} {\bibfnamefont {D.~G.}\ \bibnamefont
  {Grier}},\ }\bibfield  {title} {\bibinfo {title} {Methods of digital video
  microscopy for colloidal studies},\ }\href
  {https://doi.org/10.1006/jcis.1996.0217} {\bibfield  {journal} {\bibinfo
  {journal} {J. Coll. Int. Sci.}\ }\textbf {\bibinfo {volume} {179}},\ \bibinfo
  {pages} {298} (\bibinfo {year} {1996})}\BibitemShut {NoStop}%
\bibitem [{\citenamefont {Baumgartl}\ and\ \citenamefont
  {Bechinger}(2005)}]{Baumgartl2005}%
  \BibitemOpen
  \bibfield  {author} {\bibinfo {author} {\bibfnamefont {J.}~\bibnamefont
  {Baumgartl}}\ and\ \bibinfo {author} {\bibfnamefont {C.}~\bibnamefont
  {Bechinger}},\ }\bibfield  {title} {\bibinfo {title} {On the limits of
  digital video microscopy},\ }\href
  {https://doi.org/10.1209/epl/i2005-10107-2} {\bibfield  {journal} {\bibinfo
  {journal} {Europhys. Lett.}\ }\textbf {\bibinfo {volume} {71}},\ \bibinfo
  {pages} {487} (\bibinfo {year} {2005})}\BibitemShut {NoStop}%
\bibitem [{\citenamefont {Qiu}\ \emph {et~al.}(1990)\citenamefont {Qiu},
  \citenamefont {Wu}, \citenamefont {Xue}, \citenamefont {Pine}, \citenamefont
  {Weitz},\ and\ \citenamefont {Chaikin}}]{Qiu/etal:1990}%
  \BibitemOpen
  \bibfield  {author} {\bibinfo {author} {\bibfnamefont {X.}~\bibnamefont
  {Qiu}}, \bibinfo {author} {\bibfnamefont {X.~L.}\ \bibnamefont {Wu}},
  \bibinfo {author} {\bibfnamefont {J.~Z.}\ \bibnamefont {Xue}}, \bibinfo
  {author} {\bibfnamefont {D.~J.}\ \bibnamefont {Pine}}, \bibinfo {author}
  {\bibfnamefont {D.~A.}\ \bibnamefont {Weitz}},\ and\ \bibinfo {author}
  {\bibfnamefont {P.~M.}\ \bibnamefont {Chaikin}},\ }\bibfield  {title}
  {\bibinfo {title} {Hydrodynamic interactions in concentrated suspensions},\
  }\href {https://doi.org/10.1103/PhysRevLett.65.516} {\bibfield  {journal}
  {\bibinfo  {journal} {Phys. Rev. Lett.}\ }\textbf {\bibinfo {volume} {65}},\
  \bibinfo {pages} {516} (\bibinfo {year} {1990})}\BibitemShut {NoStop}%
\bibitem [{\citenamefont {Zahn}\ \emph {et~al.}(1997)\citenamefont {Zahn},
  \citenamefont {{M{\'e}ndez-Alcaraz}},\ and\ \citenamefont
  {Maret}}]{Zahn/etal:1997}%
  \BibitemOpen
  \bibfield  {author} {\bibinfo {author} {\bibfnamefont {K.}~\bibnamefont
  {Zahn}}, \bibinfo {author} {\bibfnamefont {J.~M.}\ \bibnamefont
  {{M{\'e}ndez-Alcaraz}}},\ and\ \bibinfo {author} {\bibfnamefont
  {G.}~\bibnamefont {Maret}},\ }\bibfield  {title} {\bibinfo {title}
  {Hydrodynamic {{Interactions May Enhance}} the {{Self}}-{{Diffusion}} of
  {{Colloidal Particles}}},\ }\href
  {https://doi.org/10.1103/PhysRevLett.79.175} {\bibfield  {journal} {\bibinfo
  {journal} {Phys. Rev. Lett.}\ }\textbf {\bibinfo {volume} {79}},\ \bibinfo
  {pages} {175} (\bibinfo {year} {1997})}\BibitemShut {NoStop}%
\bibitem [{\citenamefont {N\"agele}\ and\ \citenamefont
  {Baur}(1997)}]{Naegele/Baur:1997}%
  \BibitemOpen
  \bibfield  {author} {\bibinfo {author} {\bibfnamefont {G.}~\bibnamefont
  {N\"agele}}\ and\ \bibinfo {author} {\bibfnamefont {P.}~\bibnamefont
  {Baur}},\ }\bibfield  {title} {\bibinfo {title} {Influence of hydrodynamic
  interactions on long-time diffusion in charge-stabilized colloids},\ }\href
  {https://doi.org/10.1209/epl/i1997-00283-5} {\bibfield  {journal} {\bibinfo
  {journal} {Europhys. Lett.}\ }\textbf {\bibinfo {volume} {38}},\ \bibinfo
  {pages} {557} (\bibinfo {year} {1997})}\BibitemShut {NoStop}%
\bibitem [{\citenamefont {Rinn}\ \emph {et~al.}(1999)\citenamefont {Rinn},
  \citenamefont {Zahn}, \citenamefont {Maass},\ and\ \citenamefont
  {Maret}}]{Rinn/etal:1999}%
  \BibitemOpen
  \bibfield  {author} {\bibinfo {author} {\bibfnamefont {B.}~\bibnamefont
  {Rinn}}, \bibinfo {author} {\bibfnamefont {K.}~\bibnamefont {Zahn}}, \bibinfo
  {author} {\bibfnamefont {P.}~\bibnamefont {Maass}},\ and\ \bibinfo {author}
  {\bibfnamefont {G.}~\bibnamefont {Maret}},\ }\bibfield  {title} {\bibinfo
  {title} {Influence of hydrodynamic interactions on the dynamics of long-range
  interacting colloidal particles},\ }\href
  {https://doi.org/10.1209/epl/i1999-00297-5} {\bibfield  {journal} {\bibinfo
  {journal} {Europhys. Lett. ({EPL})}\ }\textbf {\bibinfo {volume} {46}},\
  \bibinfo {pages} {537} (\bibinfo {year} {1999})}\BibitemShut {NoStop}%
\bibitem [{\citenamefont {H\"artl}\ \emph {et~al.}(2000)\citenamefont
  {H\"artl}, \citenamefont {Wagner}, \citenamefont {Beck}, \citenamefont
  {Gierschner},\ and\ \citenamefont {Hempelmann}}]{Haertl/etal:2000}%
  \BibitemOpen
  \bibfield  {author} {\bibinfo {author} {\bibfnamefont {W.}~\bibnamefont
  {H\"artl}}, \bibinfo {author} {\bibfnamefont {J.}~\bibnamefont {Wagner}},
  \bibinfo {author} {\bibfnamefont {C.}~\bibnamefont {Beck}}, \bibinfo {author}
  {\bibfnamefont {F.}~\bibnamefont {Gierschner}},\ and\ \bibinfo {author}
  {\bibfnamefont {R.}~\bibnamefont {Hempelmann}},\ }\bibfield  {title}
  {\bibinfo {title} {Self-diffusion and hydrodynamic interactions in highly
  charged colloids},\ }\href {https://doi.org/10.1088/0953-8984/12/8A/337}
  {\bibfield  {journal} {\bibinfo  {journal} {J. Phys.: Condens. Matter}\
  }\textbf {\bibinfo {volume} {12}},\ \bibinfo {pages} {287} (\bibinfo {year}
  {2000})}\BibitemShut {NoStop}%
\bibitem [{\citenamefont {Riese}\ \emph {et~al.}(2000)\citenamefont {Riese},
  \citenamefont {Wegdam}, \citenamefont {Vos}, \citenamefont {Sprik},
  \citenamefont {Fenistein}, \citenamefont {Bongaerts},\ and\ \citenamefont
  {Gr\"ubel}}]{Riese/etal:2000}%
  \BibitemOpen
  \bibfield  {author} {\bibinfo {author} {\bibfnamefont {D.~O.}\ \bibnamefont
  {Riese}}, \bibinfo {author} {\bibfnamefont {G.~H.}\ \bibnamefont {Wegdam}},
  \bibinfo {author} {\bibfnamefont {W.~L.}\ \bibnamefont {Vos}}, \bibinfo
  {author} {\bibfnamefont {R.}~\bibnamefont {Sprik}}, \bibinfo {author}
  {\bibfnamefont {D.}~\bibnamefont {Fenistein}}, \bibinfo {author}
  {\bibfnamefont {J.~H.~H.}\ \bibnamefont {Bongaerts}},\ and\ \bibinfo {author}
  {\bibfnamefont {G.}~\bibnamefont {Gr\"ubel}},\ }\bibfield  {title} {\bibinfo
  {title} {Effective screening of hydrodynamic interactions in charged
  colloidal suspensions},\ }\href {https://doi.org/10.1103/PhysRevLett.85.5460}
  {\bibfield  {journal} {\bibinfo  {journal} {Phys. Rev. Lett.}\ }\textbf
  {\bibinfo {volume} {85}},\ \bibinfo {pages} {5460} (\bibinfo {year}
  {2000})}\BibitemShut {NoStop}%
\bibitem [{\citenamefont {Santana-Solano}\ and\ \citenamefont
  {Arauz-Lara}(2001)}]{Santana/etal:2001}%
  \BibitemOpen
  \bibfield  {author} {\bibinfo {author} {\bibfnamefont {J.}~\bibnamefont
  {Santana-Solano}}\ and\ \bibinfo {author} {\bibfnamefont {J.~L.}\
  \bibnamefont {Arauz-Lara}},\ }\bibfield  {title} {\bibinfo {title}
  {Hydrodynamic interactions in quasi-two-dimensional colloidal suspensions},\
  }\href {https://doi.org/10.1103/PhysRevLett.87.038302} {\bibfield  {journal}
  {\bibinfo  {journal} {Phys. Rev. Lett.}\ }\textbf {\bibinfo {volume} {87}},\
  \bibinfo {pages} {038302} (\bibinfo {year} {2001})}\BibitemShut {NoStop}%
\bibitem [{\citenamefont {Banchio}\ \emph {et~al.}(2006)\citenamefont
  {Banchio}, \citenamefont {Gapinski}, \citenamefont {Patkowski}, \citenamefont
  {H\"au\ss{}ler}, \citenamefont {Fluerasu}, \citenamefont {Sacanna},
  \citenamefont {Holmqvist}, \citenamefont {Meier}, \citenamefont {Lettinga},\
  and\ \citenamefont {N\"agele}}]{Banchio/etal:2006}%
  \BibitemOpen
  \bibfield  {author} {\bibinfo {author} {\bibfnamefont {A.~J.}\ \bibnamefont
  {Banchio}}, \bibinfo {author} {\bibfnamefont {J.}~\bibnamefont {Gapinski}},
  \bibinfo {author} {\bibfnamefont {A.}~\bibnamefont {Patkowski}}, \bibinfo
  {author} {\bibfnamefont {W.}~\bibnamefont {H\"au\ss{}ler}}, \bibinfo {author}
  {\bibfnamefont {A.}~\bibnamefont {Fluerasu}}, \bibinfo {author}
  {\bibfnamefont {S.}~\bibnamefont {Sacanna}}, \bibinfo {author} {\bibfnamefont
  {P.}~\bibnamefont {Holmqvist}}, \bibinfo {author} {\bibfnamefont
  {G.}~\bibnamefont {Meier}}, \bibinfo {author} {\bibfnamefont {M.~P.}\
  \bibnamefont {Lettinga}},\ and\ \bibinfo {author} {\bibfnamefont
  {G.}~\bibnamefont {N\"agele}},\ }\bibfield  {title} {\bibinfo {title}
  {Many-body hydrodynamic interactions in charge-stabilized suspensions},\
  }\href {https://doi.org/10.1103/PhysRevLett.96.138303} {\bibfield  {journal}
  {\bibinfo  {journal} {Phys. Rev. Lett.}\ }\textbf {\bibinfo {volume} {96}},\
  \bibinfo {pages} {138303} (\bibinfo {year} {2006})}\BibitemShut {NoStop}%
\bibitem [{\citenamefont {Segr\`e}\ \emph {et~al.}(1997)\citenamefont
  {Segr\`e}, \citenamefont {Herbolzheimer},\ and\ \citenamefont
  {Chaikin}}]{Segre/etal:1997}%
  \BibitemOpen
  \bibfield  {author} {\bibinfo {author} {\bibfnamefont {P.~N.}\ \bibnamefont
  {Segr\`e}}, \bibinfo {author} {\bibfnamefont {E.}~\bibnamefont
  {Herbolzheimer}},\ and\ \bibinfo {author} {\bibfnamefont {P.~M.}\
  \bibnamefont {Chaikin}},\ }\bibfield  {title} {\bibinfo {title} {Long-range
  correlations in sedimentation},\ }\href
  {https://doi.org/10.1103/PhysRevLett.79.2574} {\bibfield  {journal} {\bibinfo
   {journal} {Phys. Rev. Lett.}\ }\textbf {\bibinfo {volume} {79}},\ \bibinfo
  {pages} {2574} (\bibinfo {year} {1997})}\BibitemShut {NoStop}%
\bibitem [{\citenamefont {Segr\'e}\ \emph {et~al.}(2001)\citenamefont
  {Segr\'e}, \citenamefont {Liu}, \citenamefont {Umbanhowar},\ and\
  \citenamefont {Weitz}}]{Segre/etal:2001}%
  \BibitemOpen
  \bibfield  {author} {\bibinfo {author} {\bibfnamefont {P.~N.}\ \bibnamefont
  {Segr\'e}}, \bibinfo {author} {\bibfnamefont {F.}~\bibnamefont {Liu}},
  \bibinfo {author} {\bibfnamefont {P.}~\bibnamefont {Umbanhowar}},\ and\
  \bibinfo {author} {\bibfnamefont {D.~A.}\ \bibnamefont {Weitz}},\ }\bibfield
  {title} {\bibinfo {title} {An effective gravitational temperature for
  sedimentation},\ }\href {https://doi-org.sire.ub.edu/10.1038/35054518}
  {\bibfield  {journal} {\bibinfo  {journal} {Nature}\ }\textbf {\bibinfo
  {volume} {409}},\ \bibinfo {pages} {594} (\bibinfo {year}
  {2001})}\BibitemShut {NoStop}%
\bibitem [{\citenamefont {Padding}\ and\ \citenamefont
  {Louis}(2004)}]{Padding/Louis:2004}%
  \BibitemOpen
  \bibfield  {author} {\bibinfo {author} {\bibfnamefont {J.~T.}\ \bibnamefont
  {Padding}}\ and\ \bibinfo {author} {\bibfnamefont {A.~A.}\ \bibnamefont
  {Louis}},\ }\bibfield  {title} {\bibinfo {title} {Hydrodynamic and {Brownian}
  fluctuations in sedimenting suspensions},\ }\href
  {https://doi.org/10.1103/PhysRevLett.93.220601} {\bibfield  {journal}
  {\bibinfo  {journal} {Phys. Rev. Lett.}\ }\textbf {\bibinfo {volume} {93}},\
  \bibinfo {pages} {220601} (\bibinfo {year} {2004})}\BibitemShut {NoStop}%
\bibitem [{\citenamefont {Cui}\ \emph {et~al.}(2002)\citenamefont {Cui},
  \citenamefont {Diamant},\ and\ \citenamefont {Lin}}]{Cui/etal:2002}%
  \BibitemOpen
  \bibfield  {author} {\bibinfo {author} {\bibfnamefont {B.}~\bibnamefont
  {Cui}}, \bibinfo {author} {\bibfnamefont {H.}~\bibnamefont {Diamant}},\ and\
  \bibinfo {author} {\bibfnamefont {B.}~\bibnamefont {Lin}},\ }\bibfield
  {title} {\bibinfo {title} {Screened hydrodynamic interaction in a narrow
  channel},\ }\href {https://doi.org/10.1103/PhysRevLett.89.188302} {\bibfield
  {journal} {\bibinfo  {journal} {Phys. Rev. Lett.}\ }\textbf {\bibinfo
  {volume} {89}},\ \bibinfo {pages} {188302} (\bibinfo {year}
  {2002})}\BibitemShut {NoStop}%
\bibitem [{\citenamefont {Cui}\ \emph {et~al.}(2004)\citenamefont {Cui},
  \citenamefont {Diamant}, \citenamefont {Lin},\ and\ \citenamefont
  {Rice}}]{Cui/etal:2004}%
  \BibitemOpen
  \bibfield  {author} {\bibinfo {author} {\bibfnamefont {B.}~\bibnamefont
  {Cui}}, \bibinfo {author} {\bibfnamefont {H.}~\bibnamefont {Diamant}},
  \bibinfo {author} {\bibfnamefont {B.}~\bibnamefont {Lin}},\ and\ \bibinfo
  {author} {\bibfnamefont {S.~A.}\ \bibnamefont {Rice}},\ }\bibfield  {title}
  {\bibinfo {title} {Anomalous hydrodynamic interaction in a
  quasi-two-dimensional suspension},\ }\href
  {https://doi.org/10.1103/PhysRevLett.92.258301} {\bibfield  {journal}
  {\bibinfo  {journal} {Phys. Rev. Lett.}\ }\textbf {\bibinfo {volume} {92}},\
  \bibinfo {pages} {258301} (\bibinfo {year} {2004})}\BibitemShut {NoStop}%
\bibitem [{\citenamefont {Xu}\ \emph {et~al.}(2005)\citenamefont {Xu},
  \citenamefont {Rice}, \citenamefont {Lin},\ and\ \citenamefont
  {Diamant}}]{Xu/etal:2005}%
  \BibitemOpen
  \bibfield  {author} {\bibinfo {author} {\bibfnamefont {X.}~\bibnamefont
  {Xu}}, \bibinfo {author} {\bibfnamefont {S.~A.}\ \bibnamefont {Rice}},
  \bibinfo {author} {\bibfnamefont {B.}~\bibnamefont {Lin}},\ and\ \bibinfo
  {author} {\bibfnamefont {H.}~\bibnamefont {Diamant}},\ }\bibfield  {title}
  {\bibinfo {title} {Influence of hydrodynamic coupling on pair diffusion in a
  quasi-one-dimensional colloid system},\ }\href
  {https://doi.org/10.1103/PhysRevLett.95.158301} {\bibfield  {journal}
  {\bibinfo  {journal} {Phys. Rev. Lett.}\ }\textbf {\bibinfo {volume} {95}},\
  \bibinfo {pages} {158301} (\bibinfo {year} {2005})}\BibitemShut {NoStop}%
\bibitem [{\citenamefont {Grzybowski}\ \emph {et~al.}(2000)\citenamefont
  {Grzybowski}, \citenamefont {Stone},\ and\ \citenamefont
  {Whitesides}}]{Grzybowski/etal:2000}%
  \BibitemOpen
  \bibfield  {author} {\bibinfo {author} {\bibfnamefont {B.~A.}\ \bibnamefont
  {Grzybowski}}, \bibinfo {author} {\bibfnamefont {H.~A.}\ \bibnamefont
  {Stone}},\ and\ \bibinfo {author} {\bibfnamefont {G.~M.}\ \bibnamefont
  {Whitesides}},\ }\bibfield  {title} {\bibinfo {title} {Dynamic self-assembly
  of magnetized, millimetre-sized objects rotating at a liquid-air interface},\
  }\href {https://doi.org/10.1038/35016528} {\bibfield  {journal} {\bibinfo
  {journal} {Nature}\ }\textbf {\bibinfo {volume} {405}},\ \bibinfo {pages}
  {1033} (\bibinfo {year} {2000})}\BibitemShut {NoStop}%
\bibitem [{\citenamefont {Lenz}\ \emph {et~al.}(2003)\citenamefont {Lenz},
  \citenamefont {Joanny}, \citenamefont {J\"ulicher},\ and\ \citenamefont
  {Prost}}]{Lenz/etal:2003}%
  \BibitemOpen
  \bibfield  {author} {\bibinfo {author} {\bibfnamefont {P.}~\bibnamefont
  {Lenz}}, \bibinfo {author} {\bibfnamefont {J.-F. m.~c.}\ \bibnamefont
  {Joanny}}, \bibinfo {author} {\bibfnamefont {F.}~\bibnamefont {J\"ulicher}},\
  and\ \bibinfo {author} {\bibfnamefont {J.}~\bibnamefont {Prost}},\ }\bibfield
   {title} {\bibinfo {title} {Membranes with rotating motors},\ }\href
  {https://doi.org/10.1103/PhysRevLett.91.108104} {\bibfield  {journal}
  {\bibinfo  {journal} {Phys. Rev. Lett.}\ }\textbf {\bibinfo {volume} {91}},\
  \bibinfo {pages} {108104} (\bibinfo {year} {2003})}\BibitemShut {NoStop}%
\bibitem [{\citenamefont {Radu}\ and\ \citenamefont
  {Schilling}(2014)}]{Radu/Schilling:2014}%
  \BibitemOpen
  \bibfield  {author} {\bibinfo {author} {\bibfnamefont {M.}~\bibnamefont
  {Radu}}\ and\ \bibinfo {author} {\bibfnamefont {T.}~\bibnamefont
  {Schilling}},\ }\bibfield  {title} {\bibinfo {title} {Solvent hydrodynamics
  speed up crystal nucleation in suspensions of hard spheres},\ }\href
  {https://doi.org/10.1209/0295-5075/105/26001} {\bibfield  {journal} {\bibinfo
   {journal} {Europhys. Lett.}\ }\textbf {\bibinfo {volume} {105}},\ \bibinfo
  {pages} {26001} (\bibinfo {year} {2014})}\BibitemShut {NoStop}%
\bibitem [{\citenamefont {Tateno}\ \emph {et~al.}(2019)\citenamefont {Tateno},
  \citenamefont {Yanagishima}, \citenamefont {Russo},\ and\ \citenamefont
  {Tanaka}}]{Tateno/etal:2019}%
  \BibitemOpen
  \bibfield  {author} {\bibinfo {author} {\bibfnamefont {M.}~\bibnamefont
  {Tateno}}, \bibinfo {author} {\bibfnamefont {T.}~\bibnamefont {Yanagishima}},
  \bibinfo {author} {\bibfnamefont {J.}~\bibnamefont {Russo}},\ and\ \bibinfo
  {author} {\bibfnamefont {H.}~\bibnamefont {Tanaka}},\ }\bibfield  {title}
  {\bibinfo {title} {Influence of hydrodynamic interactions on colloidal
  crystallization},\ }\href {https://doi.org/10.1103/PhysRevLett.123.258002}
  {\bibfield  {journal} {\bibinfo  {journal} {Phys. Rev. Lett.}\ }\textbf
  {\bibinfo {volume} {123}},\ \bibinfo {pages} {258002} (\bibinfo {year}
  {2019})}\BibitemShut {NoStop}%
\bibitem [{\citenamefont {Reichert}\ and\ \citenamefont
  {Stark}(2004)}]{Reichert/Stark:2004}%
  \BibitemOpen
  \bibfield  {author} {\bibinfo {author} {\bibfnamefont {M.}~\bibnamefont
  {Reichert}}\ and\ \bibinfo {author} {\bibfnamefont {H.}~\bibnamefont
  {Stark}},\ }\bibfield  {title} {\bibinfo {title} {Circling particles and
  drafting in optical vortices},\ }\href
  {https://doi.org/10.1088/0953-8984/16/38/023} {\bibfield  {journal} {\bibinfo
   {journal} {J. Phys.: Condens. Matter}\ }\textbf {\bibinfo {volume} {16}},\
  \bibinfo {pages} {S4085} (\bibinfo {year} {2004})}\BibitemShut {NoStop}%
\bibitem [{\citenamefont {Lutz}\ \emph {et~al.}(2006)\citenamefont {Lutz},
  \citenamefont {Reichert}, \citenamefont {Stark},\ and\ \citenamefont
  {Bechinger}}]{Lutz/etal:2006}%
  \BibitemOpen
  \bibfield  {author} {\bibinfo {author} {\bibfnamefont {C.}~\bibnamefont
  {Lutz}}, \bibinfo {author} {\bibfnamefont {M.}~\bibnamefont {Reichert}},
  \bibinfo {author} {\bibfnamefont {H.}~\bibnamefont {Stark}},\ and\ \bibinfo
  {author} {\bibfnamefont {C.}~\bibnamefont {Bechinger}},\ }\bibfield  {title}
  {\bibinfo {title} {Surmounting barriers: {{The}} benefit of hydrodynamic
  interactions},\ }\href {https://doi.org/10.1209/epl/i2006-10017-9} {\bibfield
   {journal} {\bibinfo  {journal} {EPL}\ }\textbf {\bibinfo {volume} {74}},\
  \bibinfo {pages} {719} (\bibinfo {year} {2006})}\BibitemShut {NoStop}%
\bibitem [{\citenamefont {Beatus}\ \emph {et~al.}(2007)\citenamefont {Beatus},
  \citenamefont {Bar-Ziv},\ and\ \citenamefont {Tlusty}}]{Beatus/etal:2007}%
  \BibitemOpen
  \bibfield  {author} {\bibinfo {author} {\bibfnamefont {T.}~\bibnamefont
  {Beatus}}, \bibinfo {author} {\bibfnamefont {R.}~\bibnamefont {Bar-Ziv}},\
  and\ \bibinfo {author} {\bibfnamefont {T.}~\bibnamefont {Tlusty}},\
  }\bibfield  {title} {\bibinfo {title} {Anomalous microfluidic phonons induced
  by the interplay of hydrodynamic screening and incompressibility},\ }\href
  {https://doi.org/10.1103/PhysRevLett.99.124502} {\bibfield  {journal}
  {\bibinfo  {journal} {Phys. Rev. Lett.}\ }\textbf {\bibinfo {volume} {99}},\
  \bibinfo {pages} {124502} (\bibinfo {year} {2007})}\BibitemShut {NoStop}%
\bibitem [{\citenamefont {Beatus}\ \emph {et~al.}(2009)\citenamefont {Beatus},
  \citenamefont {Tlusty},\ and\ \citenamefont {Bar-Ziv}}]{Beatus/etal:2009}%
  \BibitemOpen
  \bibfield  {author} {\bibinfo {author} {\bibfnamefont {T.}~\bibnamefont
  {Beatus}}, \bibinfo {author} {\bibfnamefont {T.}~\bibnamefont {Tlusty}},\
  and\ \bibinfo {author} {\bibfnamefont {R.}~\bibnamefont {Bar-Ziv}},\
  }\bibfield  {title} {\bibinfo {title} {Burgers shock waves and sound in a
  {2D} microfluidic droplets ensemble},\ }\href
  {https://doi.org/10.1103/PhysRevLett.103.114502} {\bibfield  {journal}
  {\bibinfo  {journal} {Phys. Rev. Lett.}\ }\textbf {\bibinfo {volume} {103}},\
  \bibinfo {pages} {114502} (\bibinfo {year} {2009})}\BibitemShut {NoStop}%
\bibitem [{\citenamefont {Grimm}\ and\ \citenamefont
  {Stark}(2011)}]{Grimm/Stark:2011}%
  \BibitemOpen
  \bibfield  {author} {\bibinfo {author} {\bibfnamefont {A.}~\bibnamefont
  {Grimm}}\ and\ \bibinfo {author} {\bibfnamefont {H.}~\bibnamefont {Stark}},\
  }\bibfield  {title} {\bibinfo {title} {Hydrodynamic interactions enhance the
  performance of {B}rownian ratchets},\ }\href
  {https://doi.org/10.1039/C0SM01085E} {\bibfield  {journal} {\bibinfo
  {journal} {Soft Matter}\ }\textbf {\bibinfo {volume} {7}},\ \bibinfo {pages}
  {3219} (\bibinfo {year} {2011})}\BibitemShut {NoStop}%
\bibitem [{\citenamefont {Malgaretti}\ \emph {et~al.}(2012)\citenamefont
  {Malgaretti}, \citenamefont {Pagonabarraga},\ and\ \citenamefont
  {Frenkel}}]{Malgaretti/etal:2012}%
  \BibitemOpen
  \bibfield  {author} {\bibinfo {author} {\bibfnamefont {P.}~\bibnamefont
  {Malgaretti}}, \bibinfo {author} {\bibfnamefont {I.}~\bibnamefont
  {Pagonabarraga}},\ and\ \bibinfo {author} {\bibfnamefont {D.}~\bibnamefont
  {Frenkel}},\ }\bibfield  {title} {\bibinfo {title} {Running faster together:
  Huge speed up of thermal ratchets due to hydrodynamic coupling},\ }\href
  {https://doi.org/10.1103/PhysRevLett.109.168101} {\bibfield  {journal}
  {\bibinfo  {journal} {Phys. Rev. Lett.}\ }\textbf {\bibinfo {volume} {109}},\
  \bibinfo {pages} {168101} (\bibinfo {year} {2012})}\BibitemShut {NoStop}%
\bibitem [{\citenamefont {Dobnikar}\ \emph {et~al.}(2013)\citenamefont
  {Dobnikar}, \citenamefont {Snezhko},\ and\ \citenamefont
  {Yethiraj}}]{Dobnikar/etal:2013}%
  \BibitemOpen
  \bibfield  {author} {\bibinfo {author} {\bibfnamefont {J.}~\bibnamefont
  {Dobnikar}}, \bibinfo {author} {\bibfnamefont {A.}~\bibnamefont {Snezhko}},\
  and\ \bibinfo {author} {\bibfnamefont {A.}~\bibnamefont {Yethiraj}},\
  }\bibfield  {title} {\bibinfo {title} {Emergent colloidal dynamics in
  electromagnetic fields},\ }\href {https://doi.org/10.1039/C3SM27363F}
  {\bibfield  {journal} {\bibinfo  {journal} {Soft Matter}\ }\textbf {\bibinfo
  {volume} {9}},\ \bibinfo {pages} {3693} (\bibinfo {year} {2013})}\BibitemShut
  {NoStop}%
\bibitem [{\citenamefont {Nagar}\ and\ \citenamefont
  {Roichman}(2014{\natexlab{a}})}]{Nagar/Roichman:2014}%
  \BibitemOpen
  \bibfield  {author} {\bibinfo {author} {\bibfnamefont {H.}~\bibnamefont
  {Nagar}}\ and\ \bibinfo {author} {\bibfnamefont {Y.}~\bibnamefont
  {Roichman}},\ }\bibfield  {title} {\bibinfo {title} {Collective excitations
  of hydrodynamically coupled driven colloidal particles},\ }\href
  {https://doi.org/10.1103/PhysRevE.90.042302} {\bibfield  {journal} {\bibinfo
  {journal} {Phys. Rev. E}\ }\textbf {\bibinfo {volume} {90}},\ \bibinfo
  {pages} {042302} (\bibinfo {year} {2014}{\natexlab{a}})}\BibitemShut
  {NoStop}%
\bibitem [{\citenamefont {Mart\'inez-Pedrero}\ and\ \citenamefont
  {Tierno}(2018)}]{Martinez/Tierno:2018}%
  \BibitemOpen
  \bibfield  {author} {\bibinfo {author} {\bibfnamefont {F.}~\bibnamefont
  {Mart\'inez-Pedrero}}\ and\ \bibinfo {author} {\bibfnamefont
  {P.}~\bibnamefont {Tierno}},\ }\bibfield  {title} {\bibinfo {title} {Advances
  in colloidal manipulation and transport via hydrodynamic interactions},\
  }\href {https://doi.org/10.1016/j.jcis.2018.02.062} {\bibfield  {journal}
  {\bibinfo  {journal} {J. Colloid Interface Sci.}\ }\textbf {\bibinfo {volume}
  {519}},\ \bibinfo {pages} {296} (\bibinfo {year} {2018})}\BibitemShut
  {NoStop}%
\bibitem [{\citenamefont {Misiunas}\ and\ \citenamefont
  {Keyser}(2019)}]{Misiunas/Keyser:2019}%
  \BibitemOpen
  \bibfield  {author} {\bibinfo {author} {\bibfnamefont {K.}~\bibnamefont
  {Misiunas}}\ and\ \bibinfo {author} {\bibfnamefont {U.~F.}\ \bibnamefont
  {Keyser}},\ }\bibfield  {title} {\bibinfo {title} {Density-dependent speed-up
  of particle transport in channels},\ }\href
  {https://doi.org/10.1103/PhysRevLett.122.214501} {\bibfield  {journal}
  {\bibinfo  {journal} {Phys. Rev. Lett.}\ }\textbf {\bibinfo {volume} {122}},\
  \bibinfo {pages} {214501} (\bibinfo {year} {2019})}\BibitemShut {NoStop}%
\bibitem [{\citenamefont {Meiners}\ and\ \citenamefont
  {Quake}(1999)}]{Meiners/Quake:1999}%
  \BibitemOpen
  \bibfield  {author} {\bibinfo {author} {\bibfnamefont {J.-C.}\ \bibnamefont
  {Meiners}}\ and\ \bibinfo {author} {\bibfnamefont {S.~R.}\ \bibnamefont
  {Quake}},\ }\bibfield  {title} {\bibinfo {title} {Direct measurement of
  hydrodynamic cross correlations between two particles in an external
  potential},\ }\href {https://doi.org/10.1103/PhysRevLett.82.2211} {\bibfield
  {journal} {\bibinfo  {journal} {Phys. Rev. Lett.}\ }\textbf {\bibinfo
  {volume} {82}},\ \bibinfo {pages} {2211} (\bibinfo {year}
  {1999})}\BibitemShut {NoStop}%
\bibitem [{\citenamefont {Dufresne}\ \emph {et~al.}(2000)\citenamefont
  {Dufresne}, \citenamefont {Squires}, \citenamefont {Brenner},\ and\
  \citenamefont {Grier}}]{Dufresne/etal:2000}%
  \BibitemOpen
  \bibfield  {author} {\bibinfo {author} {\bibfnamefont {E.~R.}\ \bibnamefont
  {Dufresne}}, \bibinfo {author} {\bibfnamefont {T.~M.}\ \bibnamefont
  {Squires}}, \bibinfo {author} {\bibfnamefont {M.~P.}\ \bibnamefont
  {Brenner}},\ and\ \bibinfo {author} {\bibfnamefont {D.~G.}\ \bibnamefont
  {Grier}},\ }\bibfield  {title} {\bibinfo {title} {Hydrodynamic coupling of
  two {Brownian} spheres to a planar surface},\ }\href
  {https://doi.org/10.1103/PhysRevLett.85.3317} {\bibfield  {journal} {\bibinfo
   {journal} {Phys. Rev. Lett.}\ }\textbf {\bibinfo {volume} {85}},\ \bibinfo
  {pages} {3317} (\bibinfo {year} {2000})}\BibitemShut {NoStop}%
\bibitem [{\citenamefont {Martin}\ \emph {et~al.}(2006)\citenamefont {Martin},
  \citenamefont {Reichert}, \citenamefont {Stark},\ and\ \citenamefont
  {Gisler}}]{Martin/etal:2006}%
  \BibitemOpen
  \bibfield  {author} {\bibinfo {author} {\bibfnamefont {S.}~\bibnamefont
  {Martin}}, \bibinfo {author} {\bibfnamefont {M.}~\bibnamefont {Reichert}},
  \bibinfo {author} {\bibfnamefont {H.}~\bibnamefont {Stark}},\ and\ \bibinfo
  {author} {\bibfnamefont {T.}~\bibnamefont {Gisler}},\ }\bibfield  {title}
  {\bibinfo {title} {Direct observation of hydrodynamic rotation-translation
  coupling between two colloidal spheres},\ }\href
  {https://doi.org/10.1103/PhysRevLett.97.248301} {\bibfield  {journal}
  {\bibinfo  {journal} {Phys. Rev. Lett.}\ }\textbf {\bibinfo {volume} {97}},\
  \bibinfo {pages} {248301} (\bibinfo {year} {2006})}\BibitemShut {NoStop}%
\bibitem [{\citenamefont {Kotar}\ \emph {et~al.}(2013)\citenamefont {Kotar},
  \citenamefont {Debono}, \citenamefont {Bruot}, \citenamefont {Box},
  \citenamefont {Phillips}, \citenamefont {Simpson}, \citenamefont {Hanna},\
  and\ \citenamefont {Cicuta}}]{Kotar/etal:2013}%
  \BibitemOpen
  \bibfield  {author} {\bibinfo {author} {\bibfnamefont {J.}~\bibnamefont
  {Kotar}}, \bibinfo {author} {\bibfnamefont {L.}~\bibnamefont {Debono}},
  \bibinfo {author} {\bibfnamefont {N.}~\bibnamefont {Bruot}}, \bibinfo
  {author} {\bibfnamefont {S.}~\bibnamefont {Box}}, \bibinfo {author}
  {\bibfnamefont {D.}~\bibnamefont {Phillips}}, \bibinfo {author}
  {\bibfnamefont {S.}~\bibnamefont {Simpson}}, \bibinfo {author} {\bibfnamefont
  {S.}~\bibnamefont {Hanna}},\ and\ \bibinfo {author} {\bibfnamefont
  {P.}~\bibnamefont {Cicuta}},\ }\bibfield  {title} {\bibinfo {title} {Optimal
  hydrodynamic synchronization of colloidal rotors},\ }\href
  {https://doi.org/10.1103/PhysRevLett.111.228103} {\bibfield  {journal}
  {\bibinfo  {journal} {Phys. Rev. Lett.}\ }\textbf {\bibinfo {volume} {111}},\
  \bibinfo {pages} {228103} (\bibinfo {year} {2013})}\BibitemShut {NoStop}%
\bibitem [{\citenamefont {Kreiserman}\ \emph {et~al.}(2019)\citenamefont
  {Kreiserman}, \citenamefont {Malik},\ and\ \citenamefont
  {Kaplan}}]{Kreiserman/etal:2019}%
  \BibitemOpen
  \bibfield  {author} {\bibinfo {author} {\bibfnamefont {R.}~\bibnamefont
  {Kreiserman}}, \bibinfo {author} {\bibfnamefont {O.}~\bibnamefont {Malik}},\
  and\ \bibinfo {author} {\bibfnamefont {A.}~\bibnamefont {Kaplan}},\
  }\bibfield  {title} {\bibinfo {title} {Decoupling conservative forces and
  hydrodynamic interactions between optically trapped spheres},\ }\href
  {https://doi.org/10.1103/PhysRevE.99.012611} {\bibfield  {journal} {\bibinfo
  {journal} {Phys. Rev. E}\ }\textbf {\bibinfo {volume} {99}},\ \bibinfo
  {pages} {012611} (\bibinfo {year} {2019})}\BibitemShut {NoStop}%
\bibitem [{\citenamefont {Ladavac}\ and\ \citenamefont
  {Grier}(2005)}]{Ladavac/Grier:2005}%
  \BibitemOpen
  \bibfield  {author} {\bibinfo {author} {\bibfnamefont {K.}~\bibnamefont
  {Ladavac}}\ and\ \bibinfo {author} {\bibfnamefont {D.~G.}\ \bibnamefont
  {Grier}},\ }\bibfield  {title} {\bibinfo {title} {Colloidal hydrodynamic
  coupling in concentric optical vortices},\ }\href
  {https://doi.org/10.1209/epl/i2005-10022-6} {\bibfield  {journal} {\bibinfo
  {journal} {Europhys. Lett.}\ }\textbf {\bibinfo {volume} {70}},\ \bibinfo
  {pages} {548} (\bibinfo {year} {2005})}\BibitemShut {NoStop}%
\bibitem [{\citenamefont {Polin}\ \emph {et~al.}(2006)\citenamefont {Polin},
  \citenamefont {Grier},\ and\ \citenamefont {Quake}}]{Polin/etal:2006}%
  \BibitemOpen
  \bibfield  {author} {\bibinfo {author} {\bibfnamefont {M.}~\bibnamefont
  {Polin}}, \bibinfo {author} {\bibfnamefont {D.~G.}\ \bibnamefont {Grier}},\
  and\ \bibinfo {author} {\bibfnamefont {S.~R.}\ \bibnamefont {Quake}},\
  }\bibfield  {title} {\bibinfo {title} {Anomalous vibrational dispersion in
  holographically trapped colloidal arrays},\ }\href
  {https://doi.org/10.1103/PhysRevLett.96.088101} {\bibfield  {journal}
  {\bibinfo  {journal} {Phys. Rev. Lett.}\ }\textbf {\bibinfo {volume} {96}},\
  \bibinfo {pages} {088101} (\bibinfo {year} {2006})}\BibitemShut {NoStop}%
\bibitem [{\citenamefont {Cereceda-L\'opez}\ \emph {et~al.}(2021)\citenamefont
  {Cereceda-L\'opez}, \citenamefont {Lips}, \citenamefont {Ortiz-Ambriz},
  \citenamefont {Ryabov}, \citenamefont {Maass},\ and\ \citenamefont
  {Tierno}}]{Cerendeca-Lopez:2021}%
  \BibitemOpen
  \bibfield  {author} {\bibinfo {author} {\bibfnamefont {E.}~\bibnamefont
  {Cereceda-L\'opez}}, \bibinfo {author} {\bibfnamefont {D.}~\bibnamefont
  {Lips}}, \bibinfo {author} {\bibfnamefont {A.}~\bibnamefont {Ortiz-Ambriz}},
  \bibinfo {author} {\bibfnamefont {A.}~\bibnamefont {Ryabov}}, \bibinfo
  {author} {\bibfnamefont {P.}~\bibnamefont {Maass}},\ and\ \bibinfo {author}
  {\bibfnamefont {P.}~\bibnamefont {Tierno}},\ }\bibfield  {title} {\bibinfo
  {title} {Hydrodynamic interactions can induce jamming in flow-driven
  systems},\ }\href {https://doi.org/10.1103/PhysRevLett.127.214501} {\bibfield
   {journal} {\bibinfo  {journal} {Phys. Rev. Lett.}\ }\textbf {\bibinfo
  {volume} {127}},\ \bibinfo {pages} {214501} (\bibinfo {year}
  {2021})}\BibitemShut {NoStop}%
\bibitem [{\citenamefont {Kim}\ and\ \citenamefont
  {Karrila}(1991)}]{Kim/Karrila:1991}%
  \BibitemOpen
  \bibfield  {author} {\bibinfo {author} {\bibfnamefont {S.}~\bibnamefont
  {Kim}}\ and\ \bibinfo {author} {\bibfnamefont {S.~J.}\ \bibnamefont
  {Karrila}},\ }\href@noop {} {\emph {\bibinfo {title} {Microhydrodynamics:
  Principles and Selected Appliacations}}}\ (\bibinfo  {publisher}
  {Butterworth-Heinemann},\ \bibinfo {address} {Boston},\ \bibinfo {year}
  {1991})\BibitemShut {NoStop}%
\bibitem [{\citenamefont {Nagar}\ and\ \citenamefont
  {Roichman}(2014{\natexlab{b}})}]{nagar_collective_2014}%
  \BibitemOpen
  \bibfield  {author} {\bibinfo {author} {\bibfnamefont {H.}~\bibnamefont
  {Nagar}}\ and\ \bibinfo {author} {\bibfnamefont {Y.}~\bibnamefont
  {Roichman}},\ }\bibfield  {title} {\bibinfo {title} {Collective excitations
  of hydrodynamically coupled driven colloidal particles},\ }\href
  {https://doi.org/10.1103/PhysRevE.90.042302} {\bibfield  {journal} {\bibinfo
  {journal} {Phys. Rev. E}\ }\textbf {\bibinfo {volume} {90}},\ \bibinfo
  {pages} {042302} (\bibinfo {year} {2014}{\natexlab{b}})}\BibitemShut
  {NoStop}%
\bibitem [{\citenamefont {Sassa}\ \emph {et~al.}(2012)\citenamefont {Sassa},
  \citenamefont {Shibata}, \citenamefont {Iwashita},\ and\ \citenamefont
  {Kimura}}]{sassa_hydrodynamically_2012a}%
  \BibitemOpen
  \bibfield  {author} {\bibinfo {author} {\bibfnamefont {Y.}~\bibnamefont
  {Sassa}}, \bibinfo {author} {\bibfnamefont {S.}~\bibnamefont {Shibata}},
  \bibinfo {author} {\bibfnamefont {Y.}~\bibnamefont {Iwashita}},\ and\
  \bibinfo {author} {\bibfnamefont {Y.}~\bibnamefont {Kimura}},\ }\bibfield
  {title} {\bibinfo {title} {Hydrodynamically induced rhythmic motion of
  optically driven colloidal particles on a ring},\ }\href
  {https://doi.org/10.1103/PhysRevE.85.061402} {\bibfield  {journal} {\bibinfo
  {journal} {Phys. Rev. E}\ }\textbf {\bibinfo {volume} {85}},\ \bibinfo
  {pages} {061402} (\bibinfo {year} {2012})}\BibitemShut {NoStop}%
\bibitem [{Note1()}]{Note1}%
  \BibitemOpen
  \bibinfo {note} {Videos demonstrating the negligible particle motion in
  radial direction can be found at \protect \href
  {http://link.aps.org/supplemental/10.1103/PhysRevLett.127.214501}{doi:10.1103/PhysRevLett.127.214501}.}\BibitemShut
  {Stop}%
\bibitem [{\citenamefont {Juniper}\ \emph {et~al.}(2016)\citenamefont
  {Juniper}, \citenamefont {Straube}, \citenamefont {Aarts},\ and\
  \citenamefont {Dullens}}]{Juniper/etal:2016}%
  \BibitemOpen
  \bibfield  {author} {\bibinfo {author} {\bibfnamefont {M.~P.~N.}\
  \bibnamefont {Juniper}}, \bibinfo {author} {\bibfnamefont {A.~V.}\
  \bibnamefont {Straube}}, \bibinfo {author} {\bibfnamefont {D.~G. A.~L.}\
  \bibnamefont {Aarts}},\ and\ \bibinfo {author} {\bibfnamefont {R.~P.~A.}\
  \bibnamefont {Dullens}},\ }\bibfield  {title} {\bibinfo {title} {Colloidal
  particles driven across periodic optical-potential-energy landscapes},\
  }\href {https://doi.org/10.1103/PhysRevE.93.012608} {\bibfield  {journal}
  {\bibinfo  {journal} {Phys. Rev. E}\ }\textbf {\bibinfo {volume} {93}},\
  \bibinfo {pages} {012608} (\bibinfo {year} {2016})}\BibitemShut {NoStop}%
\bibitem [{\citenamefont {Juniper}\ \emph {et~al.}(2012)\citenamefont
  {Juniper}, \citenamefont {Besseling}, \citenamefont {Aarts},\ and\
  \citenamefont {Dullens}}]{Juniper/etal:2012}%
  \BibitemOpen
  \bibfield  {author} {\bibinfo {author} {\bibfnamefont {M.~P.~N.}\
  \bibnamefont {Juniper}}, \bibinfo {author} {\bibfnamefont {R.}~\bibnamefont
  {Besseling}}, \bibinfo {author} {\bibfnamefont {D.~G. A.~L.}\ \bibnamefont
  {Aarts}},\ and\ \bibinfo {author} {\bibfnamefont {R.~P.~A.}\ \bibnamefont
  {Dullens}},\ }\bibfield  {title} {\bibinfo {title} {Acousto-optically
  generated potential energy landscapes: Potential mapping using colloids under
  flow},\ }\href {https://doi.org/10.1364/OE.20.028707} {\bibfield  {journal}
  {\bibinfo  {journal} {Opt. Express}\ }\textbf {\bibinfo {volume} {20}},\
  \bibinfo {pages} {28707} (\bibinfo {year} {2012})}\BibitemShut {NoStop}%
\bibitem [{\citenamefont {Risken}(1985)}]{Risken:1985}%
  \BibitemOpen
  \bibfield  {author} {\bibinfo {author} {\bibfnamefont {H.}~\bibnamefont
  {Risken}},\ }\href {https://doi.org/10.1007/978-3-642-61544-3} {\emph
  {\bibinfo {title} {The {F}okker-{P}lanck {E}quation: {M}ethods of {S}olution
  and {A}pplications}}}\ (\bibinfo  {publisher} {Springer-Verlag Berlin},\
  \bibinfo {year} {1985})\BibitemShut {NoStop}%
\bibitem [{\citenamefont {Blake}(1971)}]{Blake:1971}%
  \BibitemOpen
  \bibfield  {author} {\bibinfo {author} {\bibfnamefont {J.~R.}\ \bibnamefont
  {Blake}},\ }\bibfield  {title} {\bibinfo {title} {A note on the image system
  for a stokeslet in a no-slip boundary},\ }\href
  {https://doi.org/10.1017/S0305004100049902} {\bibfield  {journal} {\bibinfo
  {journal} {Math. Proc. Camb. Philos. Soc.}\ }\textbf {\bibinfo {volume}
  {70}},\ \bibinfo {pages} {303} (\bibinfo {year} {1971})}\BibitemShut
  {NoStop}%
\bibitem [{\citenamefont {von Hansen}\ \emph {et~al.}(2011)\citenamefont {von
  Hansen}, \citenamefont {Hinczewski},\ and\ \citenamefont
  {Netz}}]{Hansen/etal:2011}%
  \BibitemOpen
  \bibfield  {author} {\bibinfo {author} {\bibfnamefont {Y.}~\bibnamefont {von
  Hansen}}, \bibinfo {author} {\bibfnamefont {M.}~\bibnamefont {Hinczewski}},\
  and\ \bibinfo {author} {\bibfnamefont {R.~R.}\ \bibnamefont {Netz}},\
  }\bibfield  {title} {\bibinfo {title} {Hydrodynamic screening near planar
  boundaries: Effects on semiflexible polymer dynamics},\ }\href
  {https://doi.org/10.1063/1.3593458} {\bibfield  {journal} {\bibinfo
  {journal} {J. Chem. Phys.}\ }\textbf {\bibinfo {volume} {134}},\ \bibinfo
  {pages} {235102} (\bibinfo {year} {2011})}\BibitemShut {NoStop}%
\bibitem [{\citenamefont {Sokolov}\ \emph {et~al.}(2011)\citenamefont
  {Sokolov}, \citenamefont {Frydel}, \citenamefont {Grier}, \citenamefont
  {Diamant},\ and\ \citenamefont {Roichman}}]{Sokolov/etal:2011}%
  \BibitemOpen
  \bibfield  {author} {\bibinfo {author} {\bibfnamefont {Y.}~\bibnamefont
  {Sokolov}}, \bibinfo {author} {\bibfnamefont {D.}~\bibnamefont {Frydel}},
  \bibinfo {author} {\bibfnamefont {D.~G.}\ \bibnamefont {Grier}}, \bibinfo
  {author} {\bibfnamefont {H.}~\bibnamefont {Diamant}},\ and\ \bibinfo {author}
  {\bibfnamefont {Y.}~\bibnamefont {Roichman}},\ }\bibfield  {title} {\bibinfo
  {title} {Hydrodynamic pair attractions between driven colloidal particles},\
  }\href {https://doi.org/10.1103/PhysRevLett.107.158302} {\bibfield  {journal}
  {\bibinfo  {journal} {Phys. Rev. Lett.}\ }\textbf {\bibinfo {volume} {107}},\
  \bibinfo {pages} {158302} (\bibinfo {year} {2011})}\BibitemShut {NoStop}%
\bibitem [{\citenamefont {Scala}\ \emph {et~al.}(2007)\citenamefont {Scala},
  \citenamefont {Voigtmann},\ and\ \citenamefont
  {De~Michele}}]{Scala/etal:2007}%
  \BibitemOpen
  \bibfield  {author} {\bibinfo {author} {\bibfnamefont {A.}~\bibnamefont
  {Scala}}, \bibinfo {author} {\bibfnamefont {T.}~\bibnamefont {Voigtmann}},\
  and\ \bibinfo {author} {\bibfnamefont {C.}~\bibnamefont {De~Michele}},\
  }\bibfield  {title} {\bibinfo {title} {Event-driven {B}rownian dynamics for
  hard spheres},\ }\href {https://doi.org/10.1063/1.2719190} {\bibfield
  {journal} {\bibinfo  {journal} {J. Chem. Phys.}\ }\textbf {\bibinfo {volume}
  {126}},\ \bibinfo {pages} {134109} (\bibinfo {year} {2007})}\BibitemShut
  {NoStop}%
\bibitem [{\citenamefont {Ryabov}\ \emph {et~al.}(2019)\citenamefont {Ryabov},
  \citenamefont {Lips},\ and\ \citenamefont {Maass}}]{Ryabov/etal:2019}%
  \BibitemOpen
  \bibfield  {author} {\bibinfo {author} {\bibfnamefont {A.}~\bibnamefont
  {Ryabov}}, \bibinfo {author} {\bibfnamefont {D.}~\bibnamefont {Lips}},\ and\
  \bibinfo {author} {\bibfnamefont {P.}~\bibnamefont {Maass}},\ }\bibfield
  {title} {\bibinfo {title} {Counterintuitive short uphill transitions in
  single-file diffusion},\ }\href {https://doi.org/10.1021/acs.jpcc.8b12081}
  {\bibfield  {journal} {\bibinfo  {journal} {J. Phys. Chem. C}\ }\textbf
  {\bibinfo {volume} {123}},\ \bibinfo {pages} {5714} (\bibinfo {year}
  {2019})}\BibitemShut {NoStop}%
\bibitem [{\citenamefont {Antonov}\ \emph
  {et~al.}(2022{\natexlab{a}})\citenamefont {Antonov}, \citenamefont {Ryabov},\
  and\ \citenamefont {Maass}}]{Antonov/etal:2022a}%
  \BibitemOpen
  \bibfield  {author} {\bibinfo {author} {\bibfnamefont {A.~P.}\ \bibnamefont
  {Antonov}}, \bibinfo {author} {\bibfnamefont {A.}~\bibnamefont {Ryabov}},\
  and\ \bibinfo {author} {\bibfnamefont {P.}~\bibnamefont {Maass}},\ }\bibfield
   {title} {\bibinfo {title} {Solitons in overdamped {Brownian} dynamics},\
  }\href {https://doi.org/10.1103/PhysRevLett.129.080601} {\bibfield  {journal}
  {\bibinfo  {journal} {Phys. Rev. Lett.}\ }\textbf {\bibinfo {volume} {129}},\
  \bibinfo {pages} {080601} (\bibinfo {year} {2022}{\natexlab{a}})}\BibitemShut
  {NoStop}%
\bibitem [{\citenamefont {Antonov}\ \emph
  {et~al.}(2022{\natexlab{b}})\citenamefont {Antonov}, \citenamefont
  {Vor{\'{a}}{\v{c}}}, \citenamefont {Ryabov},\ and\ \citenamefont
  {Maass}}]{Antonov/etal:2022b}%
  \BibitemOpen
  \bibfield  {author} {\bibinfo {author} {\bibfnamefont {A.~P.}\ \bibnamefont
  {Antonov}}, \bibinfo {author} {\bibfnamefont {D.}~\bibnamefont
  {Vor{\'{a}}{\v{c}}}}, \bibinfo {author} {\bibfnamefont {A.}~\bibnamefont
  {Ryabov}},\ and\ \bibinfo {author} {\bibfnamefont {P.}~\bibnamefont
  {Maass}},\ }\bibfield  {title} {\bibinfo {title} {Collective excitations in
  jammed states: ultrafast defect propagation and finite-size scaling},\ }\href
  {https://doi.org/10.1088/1367-2630/ac8e26} {\bibfield  {journal} {\bibinfo
  {journal} {New J. Phys.}\ }\textbf {\bibinfo {volume} {24}},\ \bibinfo
  {pages} {093020} (\bibinfo {year} {2022}{\natexlab{b}})}\BibitemShut
  {NoStop}%
\bibitem [{\citenamefont {Lips}\ \emph {et~al.}(2021)\citenamefont {Lips},
  \citenamefont {Stoop}, \citenamefont {Maass},\ and\ \citenamefont
  {Tierno}}]{Lips/etal:2021}%
  \BibitemOpen
  \bibfield  {author} {\bibinfo {author} {\bibfnamefont {D.}~\bibnamefont
  {Lips}}, \bibinfo {author} {\bibfnamefont {R.~L.}\ \bibnamefont {Stoop}},
  \bibinfo {author} {\bibfnamefont {P.}~\bibnamefont {Maass}},\ and\ \bibinfo
  {author} {\bibfnamefont {P.}~\bibnamefont {Tierno}},\ }\bibfield  {title}
  {\bibinfo {title} {Emergent colloidal currents across ordered and disordered
  landscapes},\ }\href {https://doi.org/10.1038/s42005-021-00722-0} {\bibfield
  {journal} {\bibinfo  {journal} {Commun. Phys.}\ }\textbf {\bibinfo {volume}
  {4}},\ \bibinfo {pages} {224} (\bibinfo {year} {2021})}\BibitemShut {NoStop}%
\bibitem [{\citenamefont {Durlofsky}\ \emph {et~al.}(1987)\citenamefont
  {Durlofsky}, \citenamefont {Brady},\ and\ \citenamefont
  {Bossis}}]{Durlofsky/etal:1987}%
  \BibitemOpen
  \bibfield  {author} {\bibinfo {author} {\bibfnamefont {L.}~\bibnamefont
  {Durlofsky}}, \bibinfo {author} {\bibfnamefont {J.~F.}\ \bibnamefont
  {Brady}},\ and\ \bibinfo {author} {\bibfnamefont {G.}~\bibnamefont
  {Bossis}},\ }\bibfield  {title} {\bibinfo {title} {Dynamic simulation of
  hydrodynamically interacting particles},\ }\href {https://doi.org/DOI:
  10.1017/S002211208700171X} {\bibfield  {journal} {\bibinfo  {journal} {J.
  Fluid Dyn.}\ }\textbf {\bibinfo {volume} {180}},\ \bibinfo {pages} {21}
  (\bibinfo {year} {1987})}\BibitemShut {NoStop}%
\bibitem [{\citenamefont {Swan}\ and\ \citenamefont
  {Brady}(2007)}]{Swan/Brady:2007}%
  \BibitemOpen
  \bibfield  {author} {\bibinfo {author} {\bibfnamefont {J.~W.}\ \bibnamefont
  {Swan}}\ and\ \bibinfo {author} {\bibfnamefont {J.~F.}\ \bibnamefont
  {Brady}},\ }\bibfield  {title} {\bibinfo {title} {Simulation of
  hydrodynamically interacting particles near a no-slip boundary},\ }\href
  {https://doi.org/10.1063/1.2803837} {\bibfield  {journal} {\bibinfo
  {journal} {Phys. Fluids}\ }\textbf {\bibinfo {volume} {19}},\ \bibinfo
  {pages} {113306} (\bibinfo {year} {2007})}\BibitemShut {NoStop}%
\bibitem [{\citenamefont {Howard}\ \emph {et~al.}(2019)\citenamefont {Howard},
  \citenamefont {Nikoubashman},\ and\ \citenamefont
  {Palmer}}]{Howard/etal:2019}%
  \BibitemOpen
  \bibfield  {author} {\bibinfo {author} {\bibfnamefont {M.~P.}\ \bibnamefont
  {Howard}}, \bibinfo {author} {\bibfnamefont {A.}~\bibnamefont
  {Nikoubashman}},\ and\ \bibinfo {author} {\bibfnamefont {J.~C.}\ \bibnamefont
  {Palmer}},\ }\bibfield  {title} {\bibinfo {title} {Modeling hydrodynamic
  interactions in soft materials with multiparticle collision dynamics},\
  }\href {https://doi.org/https://doi.org/10.1016/j.coche.2019.02.007}
  {\bibfield  {journal} {\bibinfo  {journal} {Curr. Opin. Chem. Eng.}\ }\textbf
  {\bibinfo {volume} {23}},\ \bibinfo {pages} {34} (\bibinfo {year}
  {2019})}\BibitemShut {NoStop}%
\bibitem [{\citenamefont {Champion}\ \emph {et~al.}(2007)\citenamefont
  {Champion}, \citenamefont {Katare},\ and\ \citenamefont
  {Mitragotri}}]{Champion2007}%
  \BibitemOpen
  \bibfield  {author} {\bibinfo {author} {\bibfnamefont {J.~A.}\ \bibnamefont
  {Champion}}, \bibinfo {author} {\bibfnamefont {Y.~K.}\ \bibnamefont
  {Katare}},\ and\ \bibinfo {author} {\bibfnamefont {S.}~\bibnamefont
  {Mitragotri}},\ }\bibfield  {title} {\bibinfo {title} {Making polymeric
  micro- and nanoparticles of complex shapes},\ }\href
  {https://doi/10.1073/pnas.0705326104} {\bibfield  {journal} {\bibinfo
  {journal} {Proc. Natl. Acad. Sci. U. S.A.}\ }\textbf {\bibinfo {volume}
  {104}},\ \bibinfo {pages} {11901} (\bibinfo {year} {2007})}\BibitemShut
  {NoStop}%
\bibitem [{\citenamefont {Rossi}\ \emph {et~al.}(2010)\citenamefont {Rossi},
  \citenamefont {Sacanna}, \citenamefont {Irvine}, \citenamefont {Chaikin},
  \citenamefont {Pine},\ and\ \citenamefont {Philipse}}]{Rossi2010}%
  \BibitemOpen
  \bibfield  {author} {\bibinfo {author} {\bibfnamefont {L.}~\bibnamefont
  {Rossi}}, \bibinfo {author} {\bibfnamefont {S.}~\bibnamefont {Sacanna}},
  \bibinfo {author} {\bibfnamefont {W.~T.~M.}\ \bibnamefont {Irvine}}, \bibinfo
  {author} {\bibfnamefont {P.~M.}\ \bibnamefont {Chaikin}}, \bibinfo {author}
  {\bibfnamefont {D.~J.}\ \bibnamefont {Pine}},\ and\ \bibinfo {author}
  {\bibfnamefont {A.~P.}\ \bibnamefont {Philipse}},\ }\bibfield  {title}
  {\bibinfo {title} {Cubic crystals from cubic colloids},\ }\href
  {https://doi.org/10.1039/C0SM01246G} {\bibfield  {journal} {\bibinfo
  {journal} {Soft Matter}\ }\textbf {\bibinfo {volume} {7}},\ \bibinfo {pages}
  {4139} (\bibinfo {year} {2010})}\BibitemShut {NoStop}%
\bibitem [{\citenamefont {Ortiz-Ambriz}\ \emph {et~al.}(2018)\citenamefont
  {Ortiz-Ambriz}, \citenamefont {Gerloff}, \citenamefont {Klapp}, \citenamefont
  {Ort\'in},\ and\ \citenamefont {Tierno}}]{Ortiz2018}%
  \BibitemOpen
  \bibfield  {author} {\bibinfo {author} {\bibfnamefont {A.}~\bibnamefont
  {Ortiz-Ambriz}}, \bibinfo {author} {\bibfnamefont {S.}~\bibnamefont
  {Gerloff}}, \bibinfo {author} {\bibfnamefont {S.~H.~L.}\ \bibnamefont
  {Klapp}}, \bibinfo {author} {\bibfnamefont {J.}~\bibnamefont {Ort\'in}},\
  and\ \bibinfo {author} {\bibfnamefont {P.}~\bibnamefont {Tierno}},\
  }\bibfield  {title} {\bibinfo {title} {Laning, thinning and thickening of
  sheared colloids in a two-dimensional {Taylor}--{Couette} geometry},\ }\href
  {https://doi.org/10.1039/C8SM00434J} {\bibfield  {journal} {\bibinfo
  {journal} {Soft Matter}\ }\textbf {\bibinfo {volume} {14}},\ \bibinfo {pages}
  {5121} (\bibinfo {year} {2018})}\BibitemShut {NoStop}%
\bibitem [{\citenamefont {Gerloff}\ \emph {et~al.}(2020)\citenamefont
  {Gerloff}, \citenamefont {Ortiz-Ambriz}, \citenamefont {Tierno},\ and\
  \citenamefont {Klapp}}]{Gerloff2020}%
  \BibitemOpen
  \bibfield  {author} {\bibinfo {author} {\bibfnamefont {S.}~\bibnamefont
  {Gerloff}}, \bibinfo {author} {\bibfnamefont {A.}~\bibnamefont
  {Ortiz-Ambriz}}, \bibinfo {author} {\bibfnamefont {P.}~\bibnamefont
  {Tierno}},\ and\ \bibinfo {author} {\bibfnamefont {S.~H.~L.}\ \bibnamefont
  {Klapp}},\ }\bibfield  {title} {\bibinfo {title} {Dynamical modes of sheared
  confined microscale matter},\ }\href {https://doi.org/10.1039/D0SM01238F}
  {\bibfield  {journal} {\bibinfo  {journal} {Soft Matter}\ }\textbf {\bibinfo
  {volume} {16}},\ \bibinfo {pages} {9423} (\bibinfo {year}
  {2020})}\BibitemShut {NoStop}%
\bibitem [{\citenamefont {Tierno}\ \emph {et~al.}(2007)\citenamefont {Tierno},
  \citenamefont {Muruganathan},\ and\ \citenamefont {Fischer}}]{Tierno2007}%
  \BibitemOpen
  \bibfield  {author} {\bibinfo {author} {\bibfnamefont {P.}~\bibnamefont
  {Tierno}}, \bibinfo {author} {\bibfnamefont {R.}~\bibnamefont
  {Muruganathan}},\ and\ \bibinfo {author} {\bibfnamefont {T.~M.}\ \bibnamefont
  {Fischer}},\ }\bibfield  {title} {\bibinfo {title} {Viscoelasticity of
  dynamically self-assembled paramagnetic colloidal clusters},\ }\href
  {https://doi.org/10.1103/PhysRevLett.98.028301} {\bibfield  {journal}
  {\bibinfo  {journal} {Phys. Rev. Lett.}\ }\textbf {\bibinfo {volume} {98}},\
  \bibinfo {pages} {028301} (\bibinfo {year} {2007})}\BibitemShut {NoStop}%
\bibitem [{\citenamefont {Martinez-Pedrero}\ \emph {et~al.}(2015)\citenamefont
  {Martinez-Pedrero}, \citenamefont {Ortiz-Ambriz}, \citenamefont
  {Pagonabarraga},\ and\ \citenamefont {Tierno}}]{Martinez2015}%
  \BibitemOpen
  \bibfield  {author} {\bibinfo {author} {\bibfnamefont {F.}~\bibnamefont
  {Martinez-Pedrero}}, \bibinfo {author} {\bibfnamefont {A.}~\bibnamefont
  {Ortiz-Ambriz}}, \bibinfo {author} {\bibfnamefont {I.}~\bibnamefont
  {Pagonabarraga}},\ and\ \bibinfo {author} {\bibfnamefont {P.}~\bibnamefont
  {Tierno}},\ }\bibfield  {title} {\bibinfo {title} {Colloidal microworms
  propelling via a cooperative hydrodynamic conveyor belt},\ }\href
  {https://doi.org/10.1103/PhysRevLett.115.138301} {\bibfield  {journal}
  {\bibinfo  {journal} {Phys. Rev. Lett.}\ }\textbf {\bibinfo {volume} {115}},\
  \bibinfo {pages} {138301} (\bibinfo {year} {2015})}\BibitemShut {NoStop}%
\end{thebibliography}

%

\end{document}